\documentclass[iop]{emulateapj}
\pdfoutput=1

\usepackage{amsmath}
\usepackage[]{hyperref}
\usepackage{xspace}

\newcommand{\unit}[1]{\ensuremath{\, \mathrm{#1}}}
\newcommand\pccm{pc\,cm$^{-3}$\xspace}

\def\apjl{ApJ}                
\def\apjs{ApJS}               
\def\aap{A\&A}                

\shorttitle{Limits on prompt, dispersed radio pulses from gamma-ray bursts}
\shortauthors{Bannister et. al.}

\begin{document}

\title{Limits on prompt, dispersed radio pulses from gamma-ray bursts}

\author{K.~W.~Bannister\altaffilmark{1,2,3}, ,T.~Murphy\altaffilmark{4}, B.~M.~Gaensler\altaffilmark{3}}
\affil{Sydney Institute for Astronomy (SIfA), School of Physics A29, \\
The University of Sydney, NSW 2006 Australia}
\email{keith.bannister@csiro.au}

\altaffiltext{1}{CSIRO Astronomy and Space Science, PO Box 76, Epping NSW 1710, Australia (current address)}
\altaffiltext{2}{Bolton Fellow}
\altaffiltext{3}{ARC Centre of Excellence for All-sky Astrophysics (CAASTRO) }
\altaffiltext{4}{School of Information Technologies, The University of Sydney, NSW 2006, Australia}

\and
\author{J.~E.~Reynolds}
\affil{CSIRO Astronomy and Space Science, PO Box 76, Epping NSW 1710, Australia}

\slugcomment{{\sc Accepted to ApJ:} July 26, 2012}

\begin{abstract}
We have searched for prompt radio emission from nine Gamma Ray Bursts (GRBs)  with a 12~m telescope at 1.4~GHz, with a time resolution of $64~\unit{\mu s}$ to 1~s.  We detected single dispersed radio pulses with significances $>6 \sigma$ in the few minutes following two GRBs. The dispersion measures of both pulses are well in excess of the expected Galactic values, and the implied rate is incompatible with known sources of single dispersed pulses. The arrival times of both pulses also coincide with breaks in the GRB X-ray light curves. A null trial and statistical arguments rule out random fluctuations as the origin of these pulses with $>95\%$ and $\sim 97\%$ confidence, respectively, although a simple population argument supports a GRB origin with confidence of only 2\%. We caution that we cannot rule out RFI as the origin of these pulses. If the single pulses are not related to the GRBs we set an upper limit on the flux density of radio pulses emitted between 200 to 1800~s after a GRB of $1.27 w^{-1/2} \unit{Jy}$, where $6.4 \times 10^{-5} \unit{s} < w < 32 \times 10^{-3} \unit{s}$ is the pulse width. We set a limit of less than 760~Jy for long timescale ($>1$~s) variations. These limits are some of the most constraining at high time resolution and GHz frequencies in the early stages of the GRB phenomenon.
\end{abstract}


\section{Introduction}
When processing archival data from the Parkes 64~m telescope, \citet{lorimer2007bmr} detected a single  30~Jy burst with a spectral index of -4 ($S \propto \nu ^{\alpha}$) with $\sim 5 \unit{ms}$ duration at a dispersion measure (DM)of $375 \unit{pc~cm^{-3}}$. The dispersion measure was much larger than the Galactic contribution in the direction \citep{Cordes02} implying an extragalactic origin. From this detection, \citet{lorimer2007bmr} derived a brightness temperature around $\sim 10^{34}~\unit{K}$ and an event rate of  $3.8 \times 10^{-4} \unit{hr^{-1} deg^{-2}}$ (based on a sample of 1). \citet{lorimer2007bmr} proposed that these parameters were broadly compatible with a Gamma Ray Burst (GRB) origin, but noted that no mechanisms had been discussed that could produce such a burst. \citet{Keane11} also report a detection with the same telescope of a somewhat weaker burst (4~Jy) with higher DM ($745 \unit{pc~cm^{-3}}$).

To date, a GRB origin for these short-timescale, GHz radio pulses has yet to be observationally tested. Furthermore, some doubt has been cast on the astronomical origin of these bursts \citep{BurkeSpolaor11lg,Kocz12}, and the state of published theoretical mechanisms with the observed timescales and frequencies has not progressed. There are some, as yet unobserved mechanisms that can produce longer timescale radio emission at lower frequencies( $\sim 100~\unit{MHz}$) both for for the collapsar model for long GRBs \citep{Usov00, Sagiv02, Inoue04, Moortgat05} and short GRBs \citep{Lipunov96, Hansen01, Pshirkov10, Shibata11},  and scattering effects may limit the observability of short-timescale bursts in certain circumstances \citep{Macquart07,Lyubarsky08}.

Nonetheless, the rewards for detecting prompt emission from GRBs are great. Many aspects of the explosion physics can be probed by measurements of the prompt radio emission, such as the jet opening angle, and Lorentz factor \citep{Macquart07}, the density and distance of any scattering material \citep{Lyubarsky08}, and the structure of the fireball magnetic field \citep{Sagiv04}. Additionally, a short radio pulse from an extragalactic source is expected to undergo dispersion as it propagates through the intergalactic medium (IGM). Such dispersion would not only provide direct evidence for the existence of the majority of the baryons in the Universe \citep{Ginzburg73}, but, for bursts of sufficiently high redshift, would also differentiate between different models of cosmic reionization history \citep{Inoue04}.

There have been a number of unsuccessful searches for prompt emission from GRBs, although no searches were sensitive to the burst of \citet{lorimer2007bmr}. Most searches have been non-directed, in which a large fraction of the sky was monitored, with the hope of a GRB occurring somewhere in this region. Early results at 151 and 408~MHz with an integration time of 300~ms  detected some pulses within $\pm 10 \unit{min}$ of the gamma-ray trigger, but did not confidently associate any with GRBs \citep{Cortiglioni81,Inzani82}. A survey at 843~MHz, which was sensitive to pulses between 0.001~ms and 800~ms, also made no definitive detections  \citep{Amy89}, although it was not clear if any GRBs were present in the field of view during the observations. More recently, \citet{Katz03} performed an all-sky survey at 611~MHz, with a time resolution of 125~ms and a flux density detection threshold of 27~kJy. This search detected $\sim 4 \times 10^6$ bursts in 18 months, but rejected 99.9\% as RFI and identified the remaining bursts with solar activity. In a similar vein, \citet{Lazio10} detected no transients above 500~Jy for pulse widths of about 300~s at 73.8~MHz. A number of other surveys have begun but are yet to publish results \citep{Balsano98, Morales05}. The only report of automatic follow-up at radio frequencies is that performed by  \citet{Koranyi95} and \citet{Dessenne96} at 151~MHz with a time resolution of 1.5~s. Based on observations of two GRBs, \citet{Dessenne96} report upper limits on any radio emission of 16--73~Jy between 5 hrs before, and 2 hrs after the GRB. A search for the evaporation of primordial black holes at 3~GHz with a time resolution of $2\mu$s also failed to detect any radio emission \citep{Osullivan78}.

The lack of radio detections of previous surveys may be due to their low operating frequencies, low time resolution, insufficient sensitivity, and low sensitivity to the GRB rate. While low frequency observations have the advantage of a large predicted radio luminosity due to steep spectral index, which is predicted by some models of prompt radio emission (e.g.  \citep{Sagiv02}), and has been observed in one case \citep{lorimer2007bmr}, some low frequency effects make detecting short duration pulses more difficult. For example, scatter broadening, which substantially reduces the detectability of radio pulses \citep{Cordes03}, and sky temperature are worse at low frequencies. Low time resolution reduces the detectability of short duration radio pulses, including those required to avoid the brightness temperature constraint from induced scattering  \citep{Lyubarsky08}. In particular, the directed searches of \citet{Koranyi95} and \citet{Dessenne96} used a relatively low time resolution of 1.5~s. The sensitivities of most of the above surveys are low, and not approaching the flux levels required to detect the burst of \citet{lorimer2007bmr}, while more sensitive blind experiments have not had the field of view and on-sky time to obtain a GRB in-beam \citep{Wayth12, BurkeSpolaor11,Deneva09}.

In this paper we describe a survey to detect prompt radio emission from GRBs at 1.4~GHz, to test whether the bursts reported by \citet{lorimer2007bmr} and \citet{Keane11}  have a GRB origin. We used a single 12~m dish that slews automatically to the GRB coordinates, based on the gamma-ray position, and observes the position with high time resolution. Our aim was twofold: to attempt to detect any GHz radio emission within the first few minutes of the GRB, and to gain experience in automating  radio follow-up for a potential future experiment.

\section{Observations}

\subsection{Telescope Configuration}
We used a single 12~m dish antenna at the Parkes radio observatory. The visible declinations ranged between $-90\deg < \delta < +20 \deg$. The dish had a 1.4~GHz horn at the prime focus, with orthogonal linear polarizations supplying room temperature low noise amplifiers bolted onto the horn, and wrapped in a thermally insulating blanket. The  $T_{\rm sys}$ was measured at approximately 100~K and the bandwidth was 220~MHz between 1230 to 1450~MHz. We estimate the system equivalent flux density to be $3.8 \times 10^3 \unit{Jy}$.  No radiometer switching was used. Both polarizations were upper-sideband mixed down to a centre frequency of 600~MHz and digitized and channelized by a digital spectrometer based on an IBOB (Internet Break-out Board) \citep{McMahon08}. Sampling was performed at 800~MHz, resulting in sampling at the second Nyquist zone. The sampled signal of each polarization was channelized with a 1024 channel polyphase filterbank spread across 400~MHz, with an approximate frequency resolution of 390~kHz. There were approximately 560 usable channels across the 220~MHz bandwidth. The cross polarization product was not formed. The channelized output was detected,  integrated and dumped every 64 microseconds with a resolution of 8 bits per channel. The integrated spectrometer data were written to disk in real time and all processing was performed offline.

\subsection{GRB response}

A control computer was connected to the Gamma-ray Coordinate Network (GCN) via the socket distribution method  \citep{Barthelmy95}. The GCN was configured to send only notifications when the time delay between the satellite detection and socket distribution was less than 1 hour, the position of the burst was more than 10 degrees above the visible horizon, and the position error was less than 1 degree (the width of the primary beam of the telescope). The GCN was configured to send notifications from the \emph{Swift}/BAT, \emph {Fermi}, \emph{Integral} and \emph{Agile} missions, although only events from \emph{Swift}/BAT passed the filter constraints.

When a GRB alert arrived from the GCN, the telescope immediately began saving the 8-bit channelized data. It then slewed to the coordinates given in the burst alert, while  saving the raw spectra to disk. Typically the slew took two minutes, after which the telescope was ``on source''. After 30 minutes on source on the initial GRB position (the tracking position was not updated if additional position refinements were sent by the GCN), data capture was stopped. To check that the receiver system was operating correctly, immediately after the GRB capture stopped, the telescope slewed to a bright pulsar, either the Vela pulsar \citep{Large68} or  PSR B1641--45 \citep{Komesaroff73} and took an additional 10 minutes of raw data once the antenna was on source.

The telescope performed pulsar monitoring when it was not observing a GRB alert, or observing a bright pulsar for the system check. The pulsar monitoring results are not reported here.

\subsection{Processing}
Our  primary aim was to search for repeating signals, and single pulses at high  time resolution ($\sim 1$~ms). Our  secondary aim was to search for slowly varying signals at  low-time resolution ($>1$~s). For the slow search, dispersion can be ignored as the dispersion delay across the observing bandwidth is $\sim 770 \unit{ms}$ for a dispersion measure (DM) of $1000 \unit{pc~cm^{-3}}$, which is less than the time resolution.  Both timescales were affected by RFI, which can increase the number of false detections.

\subsubsection{Kurtosis-based RFI excision in the frequency domain}
\label{sec:rfi}

When a channel is affected by impulsive RFI, it typically contains more samples with large values than a channel containing Gaussian noise alone. One method for measuring an excess of large values is to compute the `excess kurtosis' ($\kappa$), which compares the number of large values in a time series against the expected number of large values given a Gaussian distribution.

To determine which of the channels contained impulsive RFI, we took the 8-bit raw data and, for each channel computed excess kurtosis in one second intervals.  If we define $x[n]$ as the $n^{\rm th}$ sample in a given interval, and $N$ as the number of samples in the interval, then the excess kurtosis is given by the equation:

\begin{equation}
\kappa = \frac{\mu_4}{\sigma^4}  - 3
\end{equation}

\noindent where $\mu_4$ is the fourth moment of the distribution, defined as

\begin{eqnarray}
\mu_4 & = & \frac{1}{N}\sum_{n=0}^{N-1}{(x[n] - \mu)^4} \\
\mu & = & \frac{1}{N}\sum_{n=0}^{N-1}{x[n]},
\end{eqnarray}

\noindent and $\sigma^4$ is the square of the variance of the distribution, given by

\begin{equation}
\sigma^4 = \left (\frac{1}{N-1} \sum_{n=0}^{N-1}(x[n] - \mu)^2 \right )^2 .
\end{equation}

The excess kurtosis is calculated for each channel in one-second bins, and displayed in an interactive program. This program plots the maximum and average kurtosis as a function of channel number. The maximum kurtosis for a channel that was not affected by RFI was $\kappa \simeq 0.6$, and the RFI affected channels had larger values of $\kappa$ (Figure \ref{fig:rfi}). An RFI channel mask was created by setting all channels with an average $\kappa$ less than a threshold to a weight of $1$, and the remaining channels to a weight of zero. The threshold could be set interactively in the program, and was set to between 0.6 and 0.8. In a subsequent step, a user-defined number of additional ``padding'' channels were set to zero on either side of a channel that had zero weight because of the threshold step. The number of additional padding channels was between 1 and 3.

The kurtosis was calculated separately for each polarization, but the same threshold and number of padding channels were used for each. This resulted in different channel masks for each polarization. The final channel mask was the logical `and' of both channel masks.

The resulting channel masked was used for both for  low-, and high-time resolution processing.

\subsubsection{Low-time resolution processing}

For low-time resolution processing, we took the mean of each channel in one-second bins. To compute a light curve, we weighted the spectra by the channel mask that we calculated according to the method in section \ref{sec:rfi}, and took the mean value across frequency. We then took the mean of the light curves of both polarizations.

The final product was a radio light curve, which we visually inspected for variability or transient emission.

\subsubsection{High-time resolution processing}
\label{sec:htr_processing}

Each channel in each polarization was offset to have zero mean and scaled to a variance of unity. The offset and scale were updated every 10 seconds. Corresponding channels in both polarizations were then added together and the result truncated to 2 bits. A periodicity search was performed in an almost identical manner to \citet{Keith10} and single-pulse search was performed identically to \citet{BurkeSpolaor11}.  Each time series was dedispersed with 1991 DM trials between 0 and $10^4\unit{~pc~cm^{-3}}$. The maximum DM was chosen to be well in excess of the expected Galactic and host contributions (probably $< 1000$~\pccm in both cases), and the predicted intergalactic medium for GRBs out to redshift $z \sim 6$ of $\ge 6000$~\pccm \citep{Ioka03,Inoue04}.

DM steps were chosen so that the smearing due to the finite DM step was at most 1.25 times the DM smearing in a frequency channel at the center frequency. No resampling for DMs larger than the diagonal DM was performed. This strategy results increasing DM steps, with small spacings at lower DMs and larger spacings at higher DMs. Time-based RFI excision was applied. Unlike the processing of \citet{Keith10}, a frequency mask (calculated in Section \ref{sec:rfi}) was applied, instead of the fast Fourier transform (FFT) based channel masking approach. An FFT based harmonic search was performed to search for periodic or pulsar-like signals, and a single-pulse search was also performed to search for bursty signals. The single-pulse search used the dedispersed time series, and applied 9 different boxcars of width $2^i$ samples ($0 \le i \le 9$) to search for pulse widths ranging from $64 \mu \rm{s}$ to 32~ms. The single-pulse candidates were grouped across DM and boxcar trials using a friend-of-friends algorithm \citep{BurkeSpolaor11}. This algorithm exploits the fact that  a broad-band pulse well above the minimum detection threshold appears in multiple DM and boxcar trials. Statistical variations near the threshold tend to appear in only a single DM and boxcar trial. Therefore, this algorithm declares a candidate only if a pulse is detected in a group of 3 or more adjacent DM and boxcar trials with each detection having a significance of $> 6 \sigma$. Groups with less than three detections were classified as Gaussian fluctuations, and groups whose peak significance was at a DM of less than 2~\pccm were classified as RFI.

Visual inspection of the resulting candidates clearly showed that our sample contained substantial impulsive RFI. In particular, we observed impulsive RFI that was detected strongly at low DMs and then disappeared at intermediate DMs, only to reappear again at higher DMs. The time difference between high-DM and low-DM `islands' of detections was occasionally as large as 2~s. The exact cause for this disappearance is not clear. The result was that the low-DM detections were classified as RFI (as they peaked near zero DM), but nearby high-DM detections were  erroneously classified as detections, as the friend-of-friends algorithm failed to join a continuous-stream of detections from the high DMs to the low DMs. To counteract the island problem, we rejected high-DM candidates that had time-coincident low-DM detection. I.e. we re-classified candidates initially produced by the pipeline as RFI if there was a  $> 6 \sigma$ detection within 3~s of the candidate and with a DM of $< 25$~pc~cm$^{-3}$. Visual inspection of the resulting candidates confirmed that all remaining RFI was excised by this additional filter, and that no believable low-DM candidates were excised.

Finally, we visually inspected plots of the candidates which were produced by the pipeline using the {\sc PSRCHIVE} package \citep{Hotan04} and our own custom tools.

\section{Results}

We describe the results of the single pulse, periodic and low-time resolution searches in this section. We note that the dispersion delay between the gamma-rays and the 1.4~GHz observing frequency, at the maximum DM of $10000 \unit{pc~cm^{-3}}$ is  $<30$~s, which is small enough to ignore in the following analysis.

\subsection{Events}

The telescope was operating between 2010 June and 2011 February, during which it responded to 16 \emph{Swift}/BAT alerts of which nine were confirmed GRBs (Table \ref{tab:events}). The remainder were instrumental effects, associated with known recurring objects, or Galactic sources such as X-ray binaries.

Of the nine GRBs, two had a duration $< 2$~s, satisfying the definition of short GRBs. The remaining seven GRBs were long GRBs. We detected no radio counterparts in the single pulse, periodic or long integrations for any short GRBs.

\begin{turnpage}
\begin{table*}
\centering
\scriptsize
\caption{\scriptsize Details of 9 GRBs observed in our experiment. The columns are: the GCN Trigger ID, the source name, the trigger date, the GCN circular that describes the detection, the GRB duration (i.e. the duration of activity in the BAT lightcurve),  the interval between the GRB detection and the 12~m telescope being `on source' ($T_{\rm on}$), the time from the GRB trigger to the arrival of a single pulse where detected ($T_{\rm pulse}$), the DM of the pulse, the DM distance ($D_{\rm DM}$) from the NE2001 electron density model of \citet{Cordes02}, the S/N of the detected pulse, the pulse width and a comment. The GCN circulars describe a number of attempts to detect host galaxies by a number of groups, but no host galaxies were conclusively identified for any GRBs listed except for GRB 100625A \citep{GCN10906}.}
\begin{tabular}{l l l l r r r r r r r r p{3cm}}
\tableline
& 		& 		& 			& 				& 		& \multicolumn{5}{c}{1.4~GHz Candidate} &  \\ \cline{7-12}
Trigger ID & Source  & Trigger Date  & Circular & Duration & $T_{\rm on}$ & Type & $T_{\rm pulse}$ & DM & $D_{\rm DM}$ & S/N & Width & Comment \\
& 	 		&		(UT)		& & (s)	&	(s)	&		& (s)  & (pc cm$^{-3}$) & (kpc) &	 & (ms) \\
\tableline
425647 & GRB 100625A & 2010-06-25 18:32:28 & \citet{GCN10884} &0.33 &  136 & - & - & - & - & - & - & Host galaxy detected with r $\sim$ 23 \\ 
426114 & GRB 100628A & 2010-06-28 08:16:40 & \citet{GCN10895} &0.1  &  160 & - & - & - & - & - & - & \\
&             &                     &                  &     &      & - & - & - & - & - & - & \\
426722 & GRB 100704A & 2010-07-04 03:35:08 & \citet{GCN10929} &280  &  187 & Single & 1076 & 194.57 & $> 48$ &  6.2 & 6 \\
432420 & GRB 100823A & 2010-08-23 17:25:35 & \citet{GCN11135} &16.9 &  141 & - & - & - & - & - & - & \\
436737 & GRB 100928A & 2010-09-28 02:19:52 & \citet{GCN11310} & 3.3  &  131 & - & - & - & - & - & - & No \emph{Swift} follow-up due to sun constraint. Sun angle: $46^{\circ}$\\
436094 & GRB 101011A & 2010-10-11 16:58:35 & \citet{GCN11331} & 50   &  105 & Single & 524  & 569.98 & $ > 35$ & 6.6 & 25 & \\
436737 & GRB 101020A & 2010-10-20 23:40:41 & \citet{GCN11357} &177  &  317 & - & - & - & - & - & - & No \emph{Swift} follow-up due to sun constraint. Sun angle: $26^{\circ}$\\
451191 & GRB 110412A & 2011-04-12 07:33:21 & \citet{GCN11922} &30   &  214 & Periodic & 0-1200  &  50.73   &  $>46$ & 6.1 & 5  & No \emph{Swift} follow-up due to moon constraint. Moon angle: $10^{\circ}$. Candidate period 524.48~ms \\
451343 & GRB 110414A & 2011-04-14 07:42:14 & \citet{GCN11931} &230  &  192 &  - & - & - & - & - & - & \\
\\
\tableline
\end{tabular}
\label{tab:events}
\pagestyle{empty}

\end{table*}
\end{turnpage}

\subsection{Single-pulse search}
\label{sec:singlepulse}

The friends-of-friends algorithm detected candidates for all GRBs we observed, and after the post-filtering step was applied, eleven candidates remained. We rejected nine of these candidates because they had relatively low DM ($<200$~\pccm) and occurred during periods of frequent, strong RFI (i.e. within $\sim 100~\unit{s}$, of the candidate).

\begin{figure*}
\centering
\includegraphics[height=0.23\textheight]{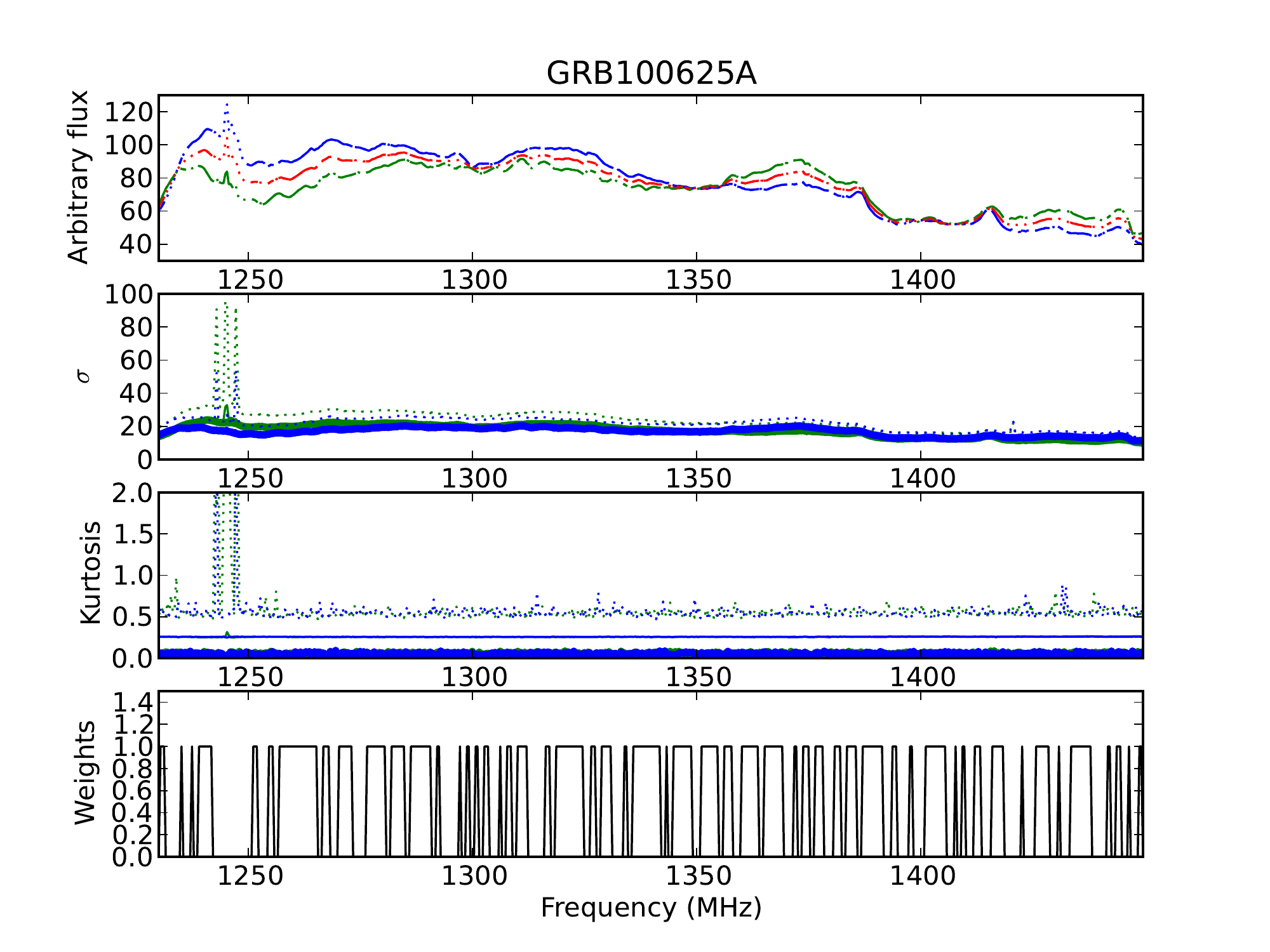}
\includegraphics[height=0.23\textheight]{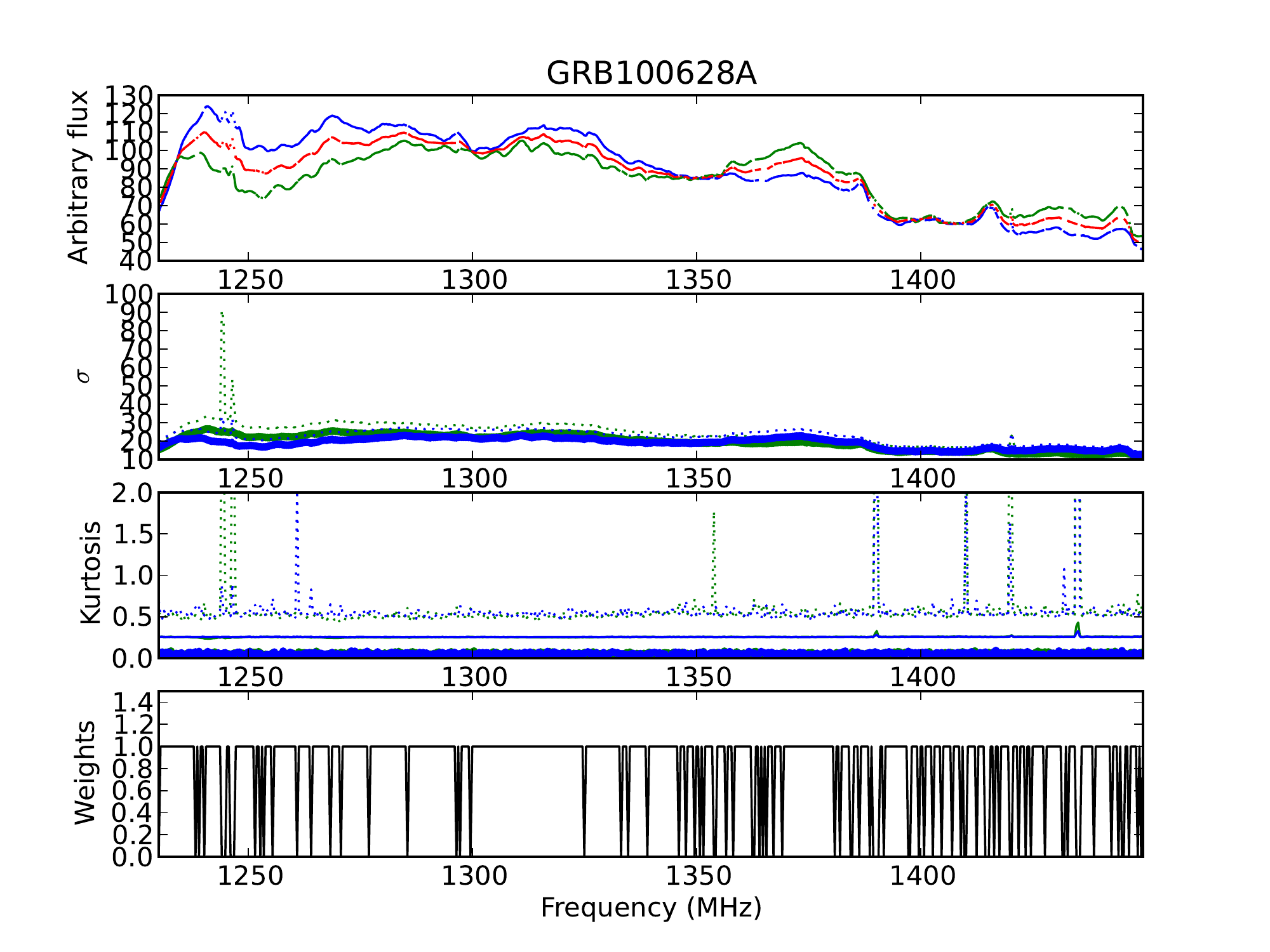}
\includegraphics[height=0.23\textheight]{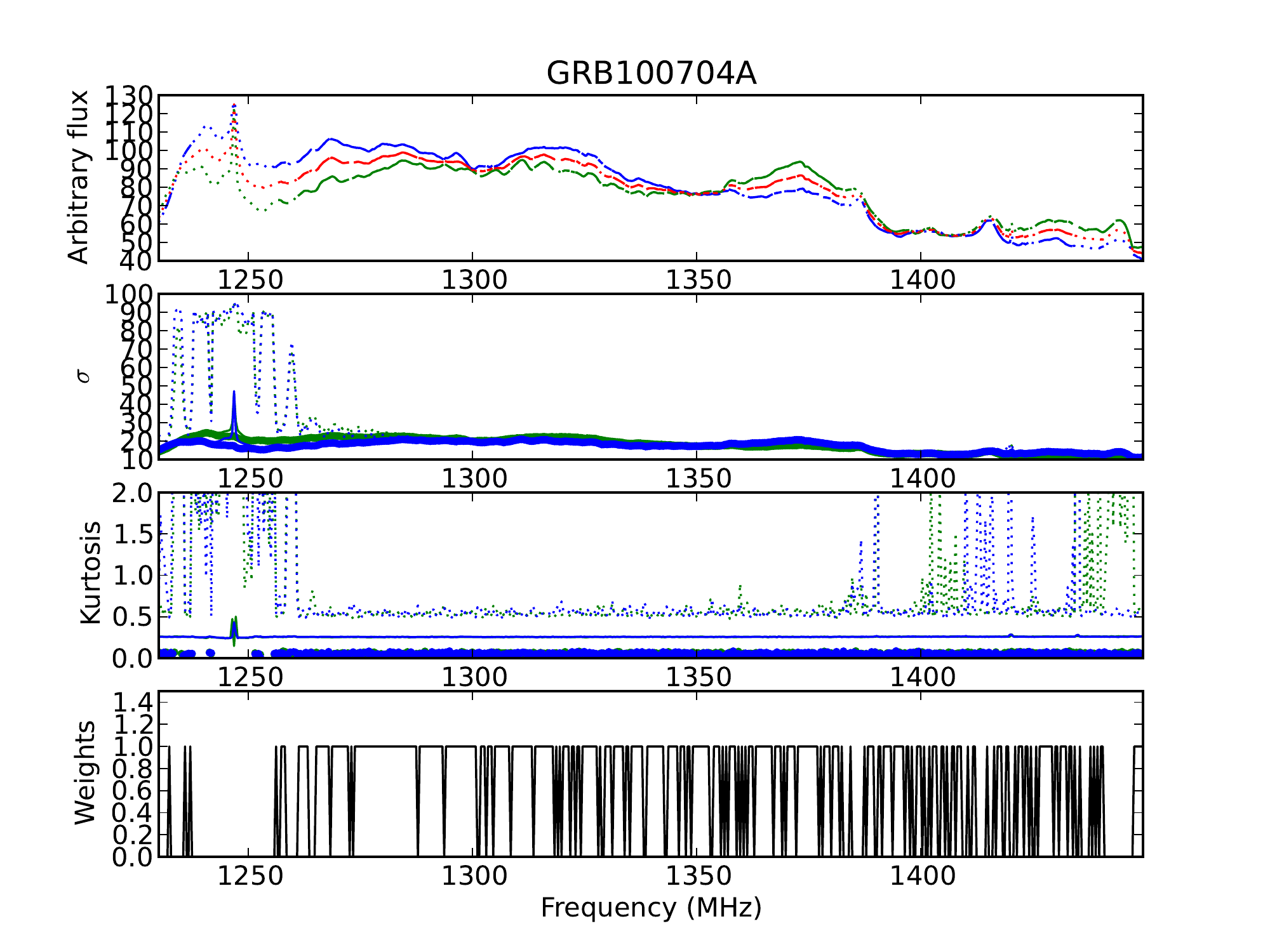}
\includegraphics[height=0.23\textheight]{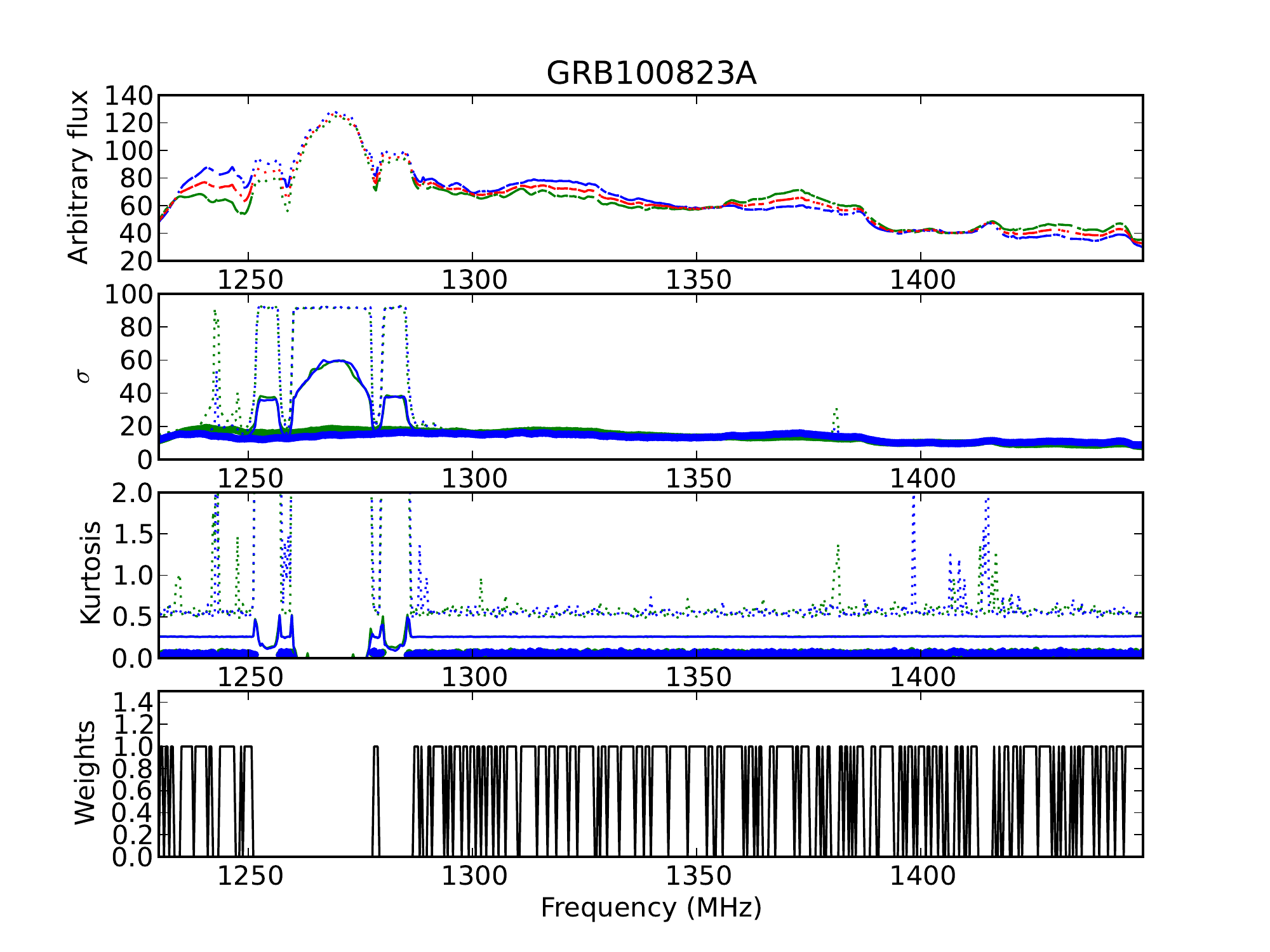}
\includegraphics[height=0.23\textheight]{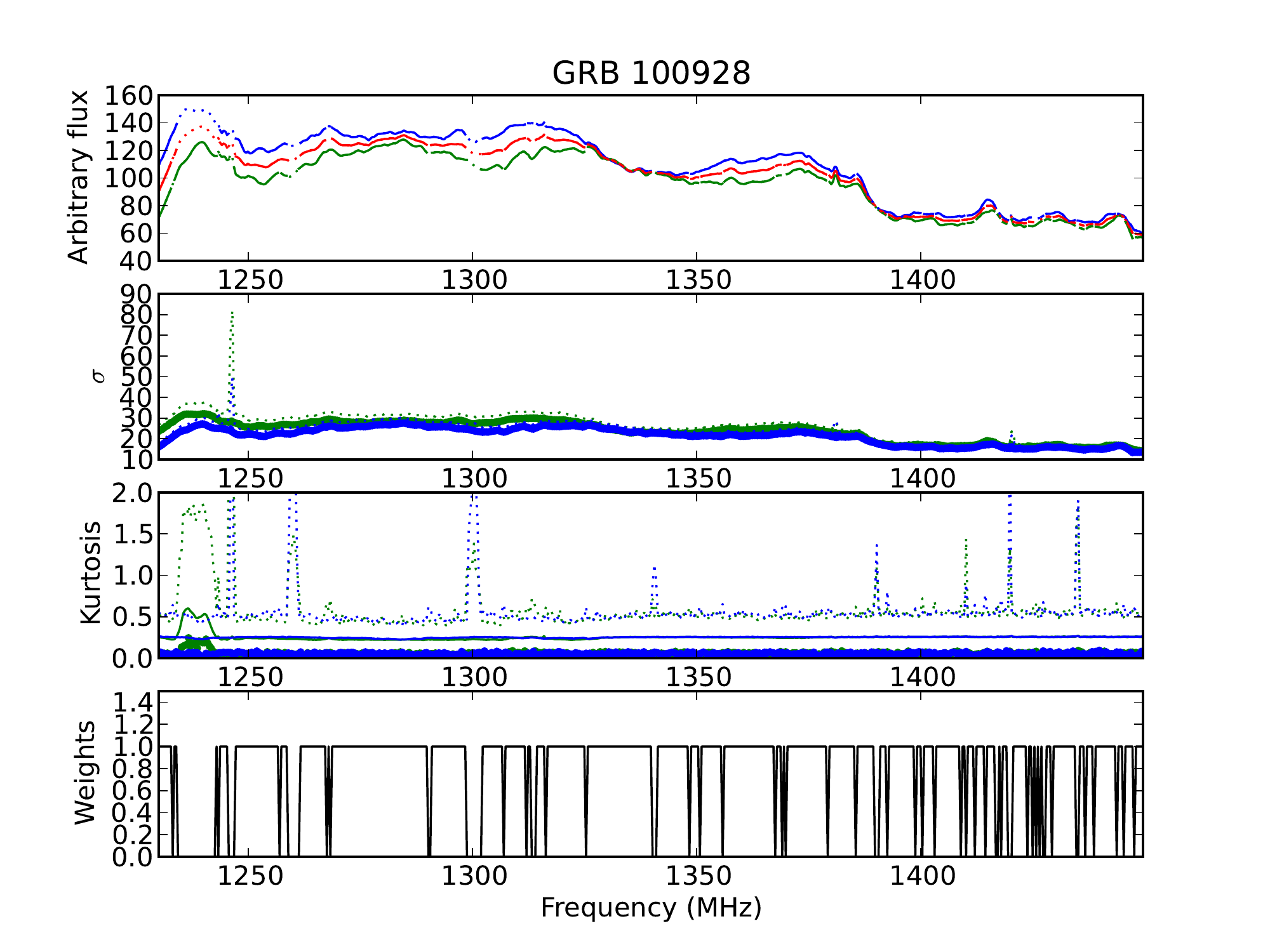}
\includegraphics[height=0.23\textheight]{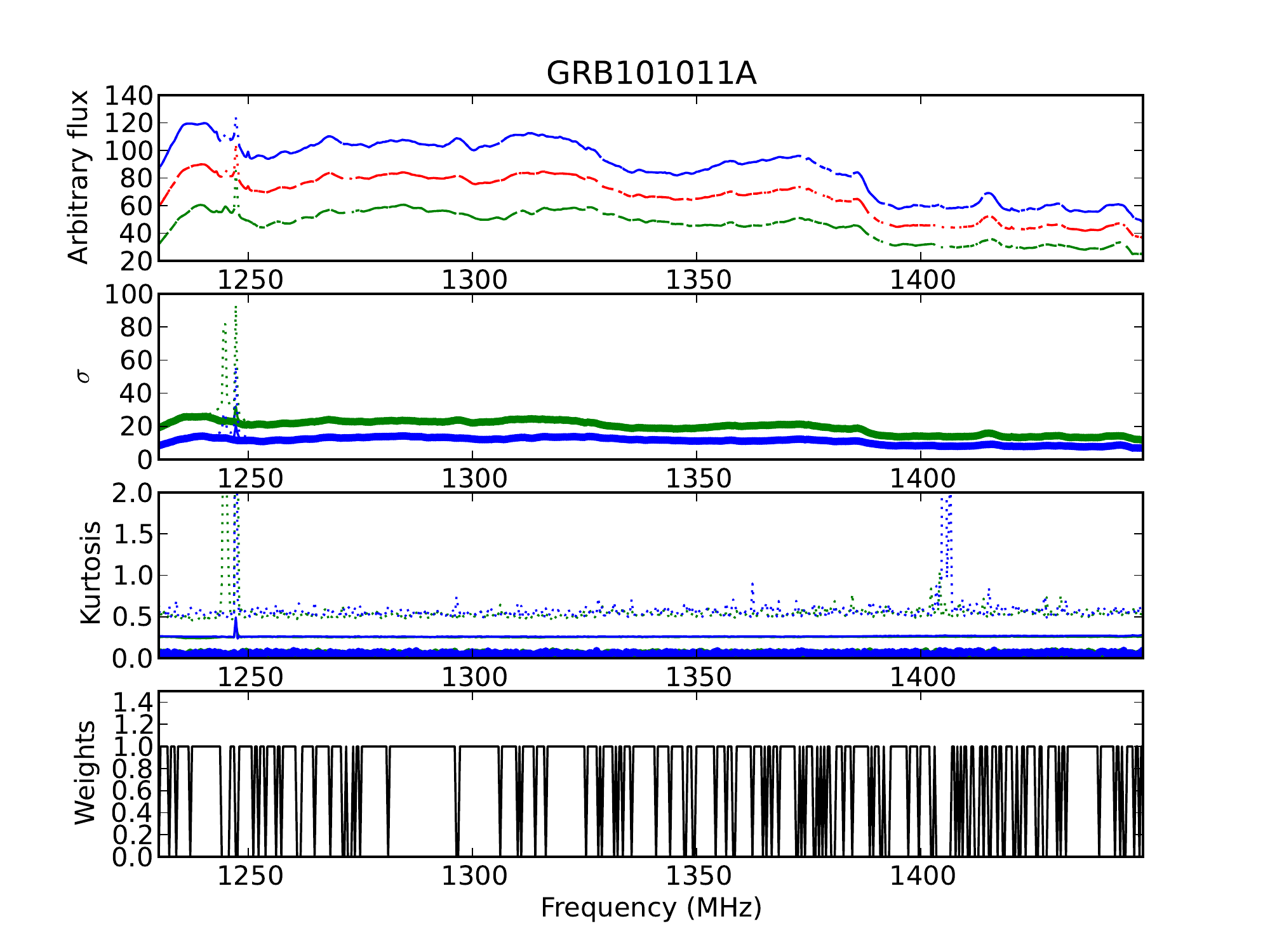}
\includegraphics[height=0.23\textheight]{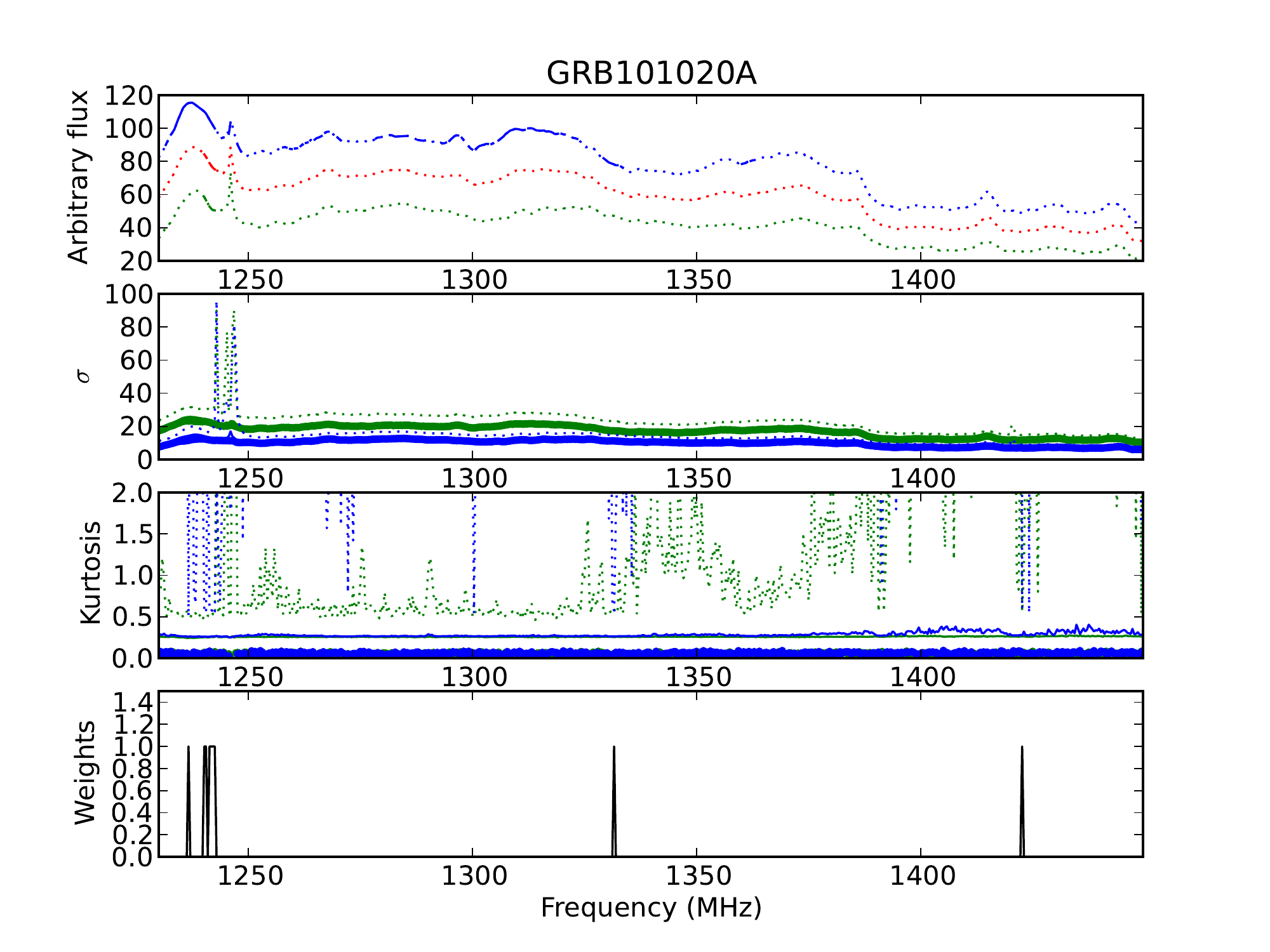}
\includegraphics[height=0.23\textheight]{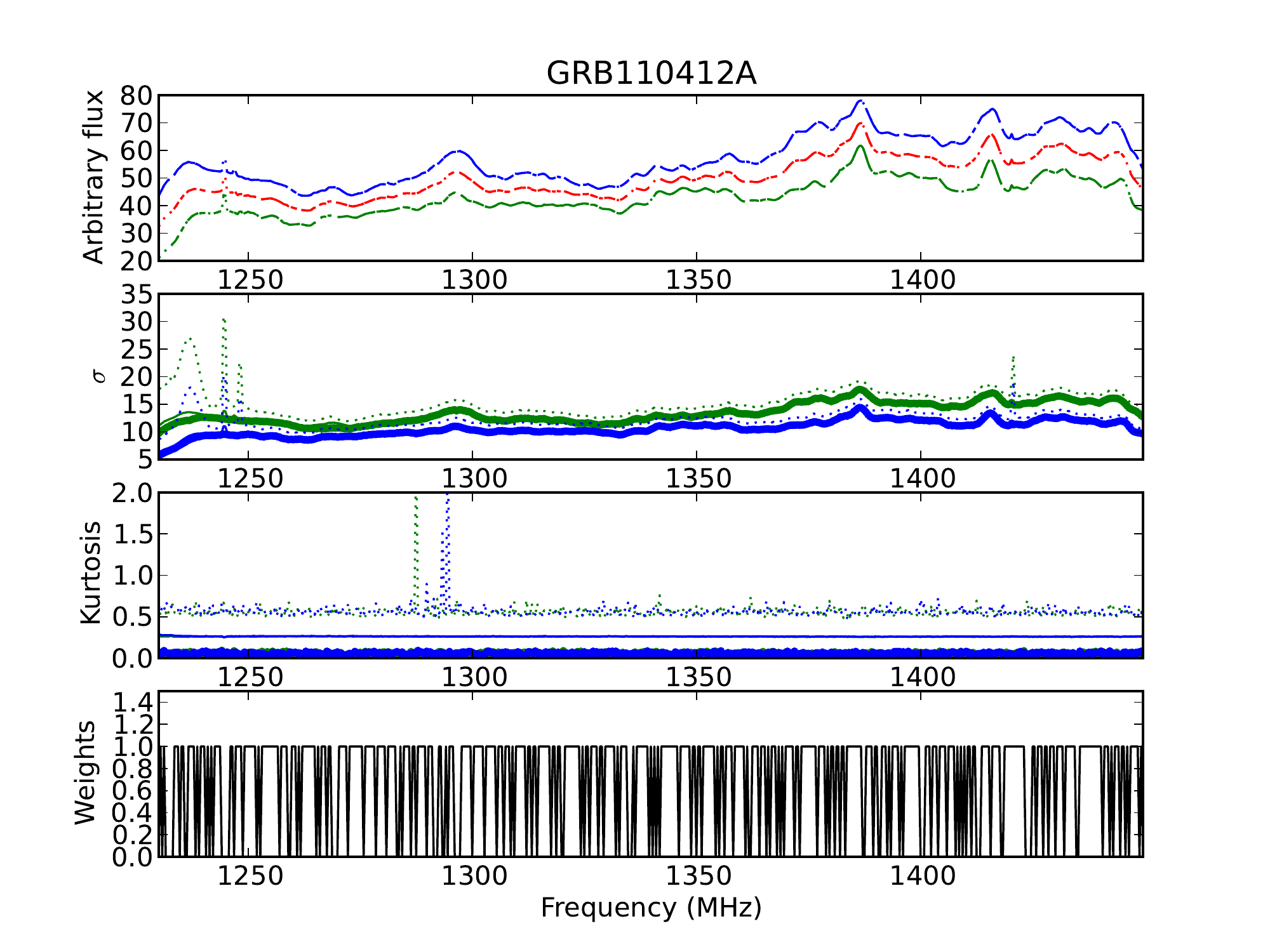}
\caption{Plots of time-averaged long-time integrations for each GRB observation. For each GRB top panel: bandpass, 2nd panel: standard deviation, 3rd panel: kurtosis, 4th panel: channel weights. Solid lines are the mean values over 30~min, and dotted lines are the maximum values over 30~min.}
\label{fig:rfi}
\end{figure*}

\begin{figure}
\centering
\includegraphics[width=\linewidth]{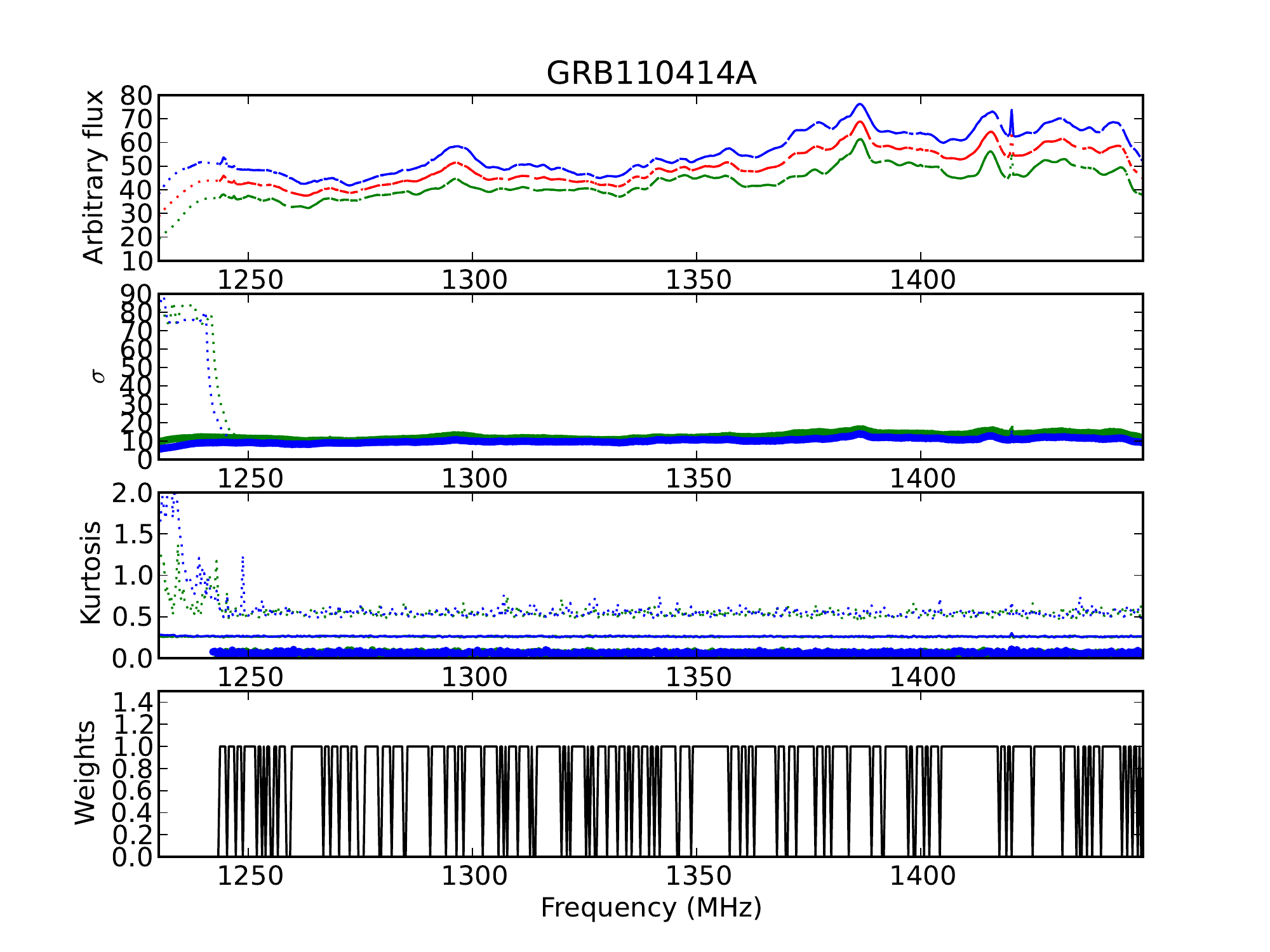}
Figure \ref{fig:rfi} \emph{continued.}
\end{figure}

\begin{figure*}
\includegraphics[width=\textwidth]{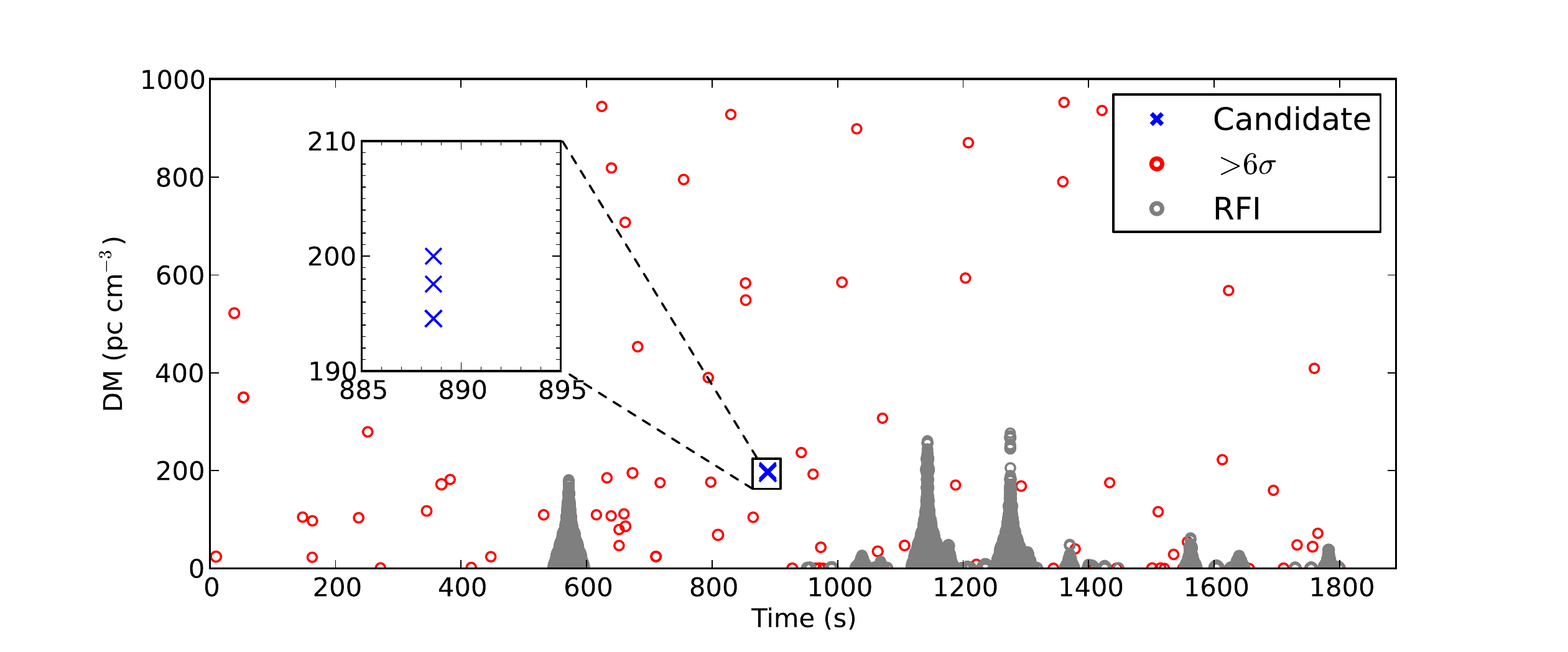}
\includegraphics[width=\textwidth]{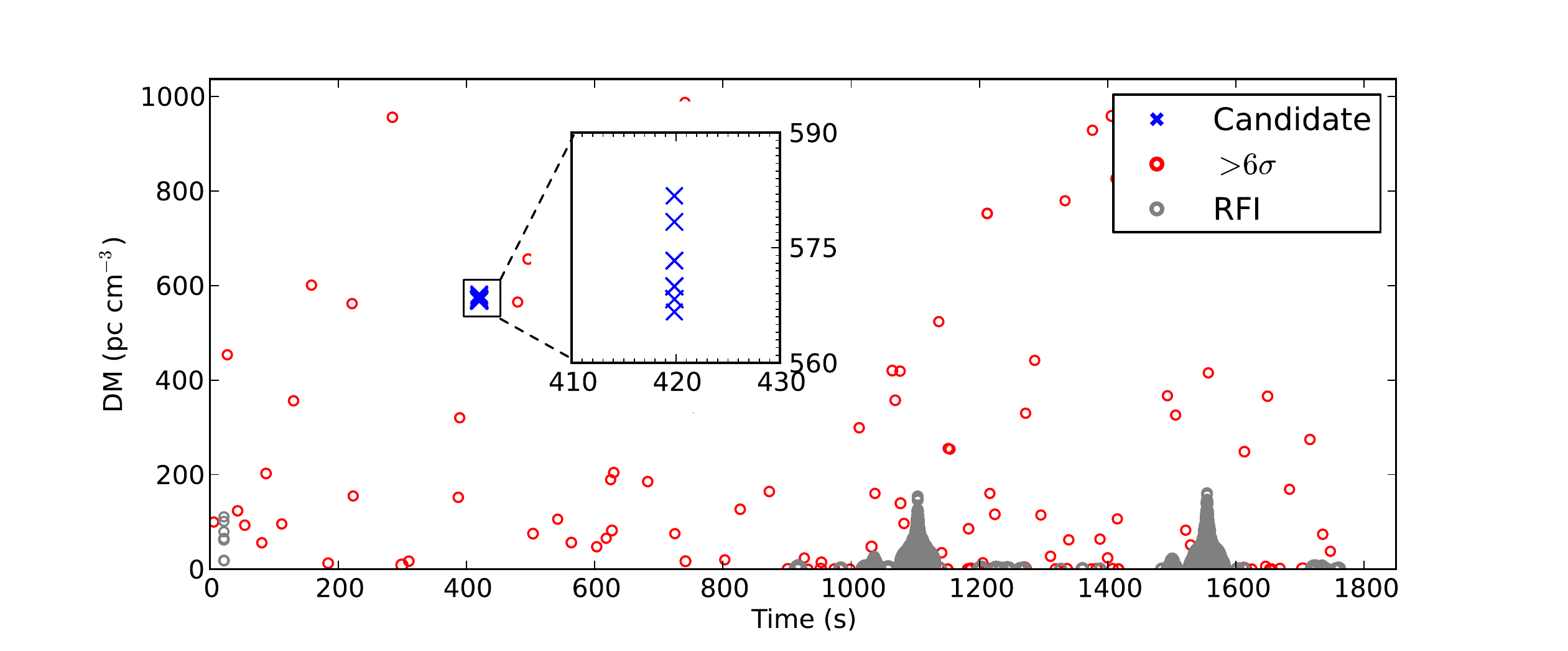}
\caption{DM vs. time for the two GRBs with single-pulse candidates. Single-pulse detections with a significance $\ge 6 \sigma$ appear as circles in this plot, with the size proportional to the S/N. The detections are color coded according to their classification by the friends-of-friends algorithm as candidates (blue), false positives (red) and RFI (grey). Top panel: GRB 100704A with a single pulse candidate 1076 seconds after the GRB at a DM of 195 pc cm$^{-3}$, with a significance of $6.2 \sigma$ and width of 6~ms. Bottom panel: GRB 101011A  with a single pulse candidate 524 seconds after the GRB at a DM of 570 pc cm$^{-3}$, with a significance of $6.6 \sigma$ and width of 25~ms. The time origin of these plots is the time that the telescope first arrived on source ($T_{\rm on}$). For clarity, DMs from 1000 pc cm$^{-3}$ are not shown.}
\label{fig:search_plots}
\end{figure*}

\begin{figure}
\centering
\includegraphics[width=\linewidth]{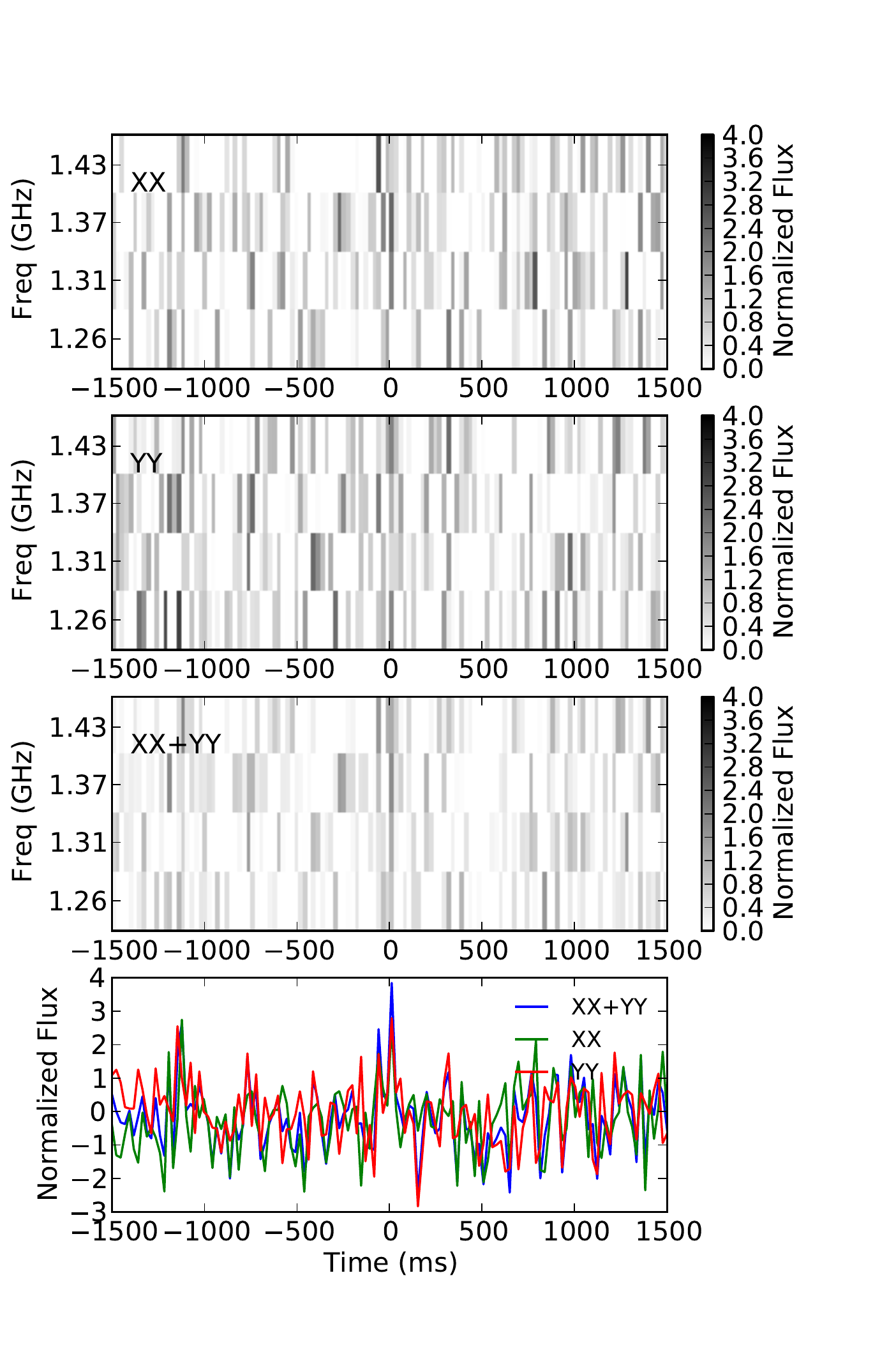}

\caption{Detail of the $6.2\sigma$ pulse detected 1076 seconds after GRB 100704A. The top three panels are the dedispersed time series of 4 frequency channels spread across the band. The pulse appears clearly in both polarizations (the top 2 panels) and in the sum of the two polarizations (third panel). The bottom panel is the time series where all the frequency channels have been summed. The origin of the time axis is the pulse arrival time ($T_{\rm pulse}$).}
\label{fig:426722_pulse}
\end{figure}

\begin{figure}
\centering
\includegraphics[width=\linewidth]{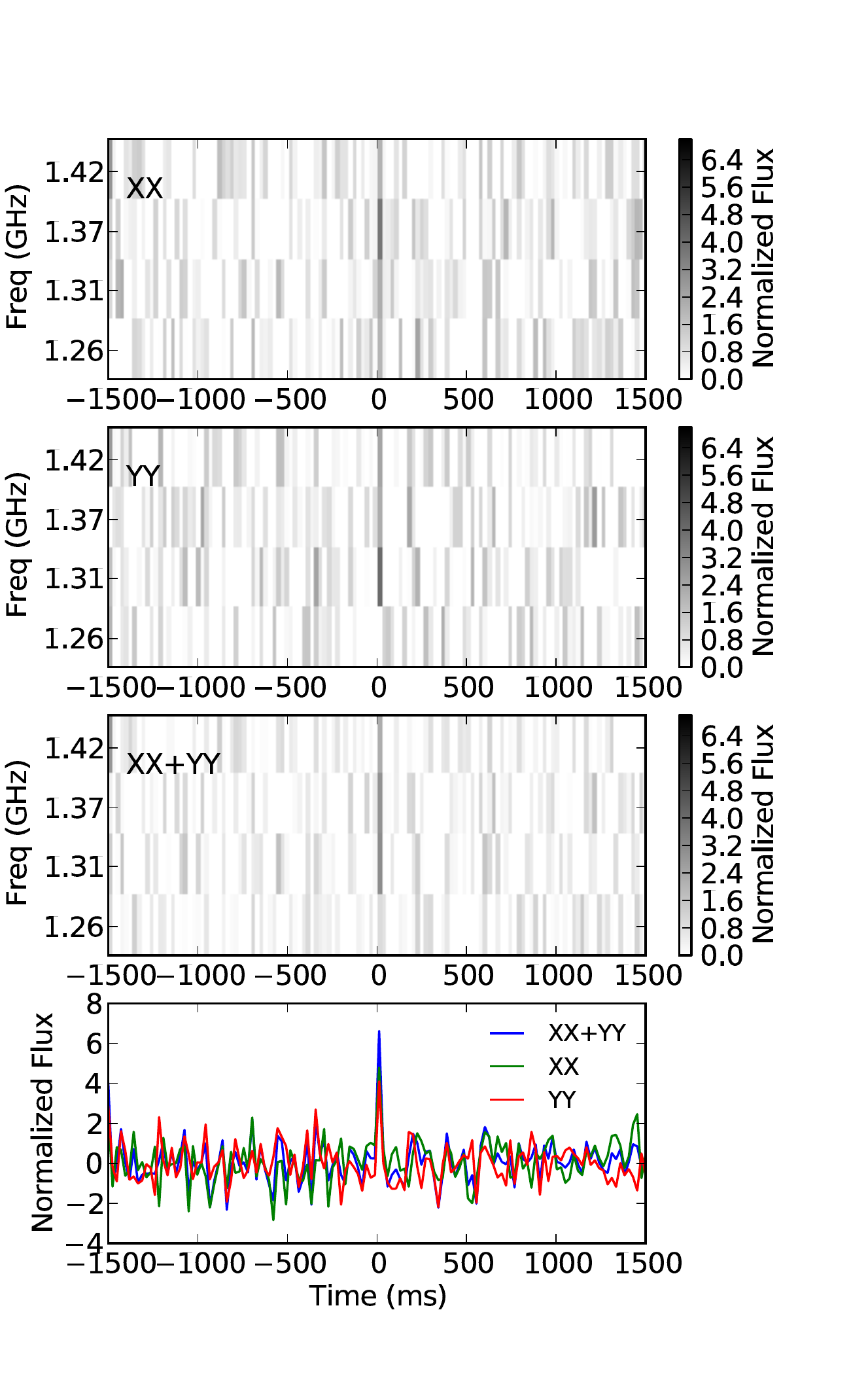}

\caption{Detail of the $6.6\sigma$ pulse detected 524 seconds after GRB 101011A. The top three panels are the dedispersed time series of 4 frequency channels spread across the band. The pulse is clearly detected across all channels, in both polarizations (the top 2 panels) and in the sum of the two polarizations (third panel). The bottom panel is the time series where all the frequency channels have been summed. The origin of the time axis is the pulse arrival time ($T_{\rm pulse}$).}
\label{fig:436094_pulse}

\end{figure}

Of the nine GRBs for which we obtained data, eight had reasonably interference-free observations (the exception being GRB 101020A, which we will not consider any further, due to substantial flagging; see Fig. \ref{fig:rfi} , 7th panel) and  two GRBs had one single radio pulse which was not associated with low-DM RFI (Figure \ref{fig:search_plots}). For GRB 100704A, a 6~ms wide single pulse was detected 1076 seconds after the GRB at a DM of 195 pc cm$^{-3}$, at a significance of $6.2 \sigma$ . This pulse is fairly indistinct in the time domain (Figure \ref{fig:426722_pulse}). For GRB 101011A, a 25~ms single pulse was detected 524 seconds after the trigger at a DM of 570~\pccm, at a significance of $6.6 \sigma$ (Figure \ref{fig:436094_pulse}).

The low significance of both detection means that fitting a spectral index is difficult. Both pulses appear to be fairly consistent with a spectral index of zero, which we will assume for the remainder of the analysis.

Both pulses have approximately the same energy in both polarizations (Figure \ref{fig:426722_pulse}, \ref{fig:436094_pulse}), arguing that they are either unpolarized, circularly polarized, or linearly polarized with a position angle approximately $45^{\circ}$ to the feed angles. A linearly polarized pulse parallel to the feed axis can be definitely ruled out by our data, as such a pulse would appear in only one polarization.  The chance of the polarization angle being aligned at $45^\circ$ to the feed is small. We therefore consider it more likely that the pulse is unpolarized, or circularly polarized.

\subsubsection{Event rate}
We detected two pulses in 4 hrs of observing. The full width, half maximum beam width of the 12~m antenna at 1.4~GHz is approximately 1.3~deg, which implies an area of $1.3 \unit{deg^2}$. If these events are not related to the GRBs and could have been discovered in a blind search, then the implied event rate is $2.9 \times 10^{-1} \unit{deg^{-2} hr^{-1} }$.

\subsubsection{Correspondence with X-ray light curves}

\begin{figure}
\centering
\includegraphics[width=\linewidth]{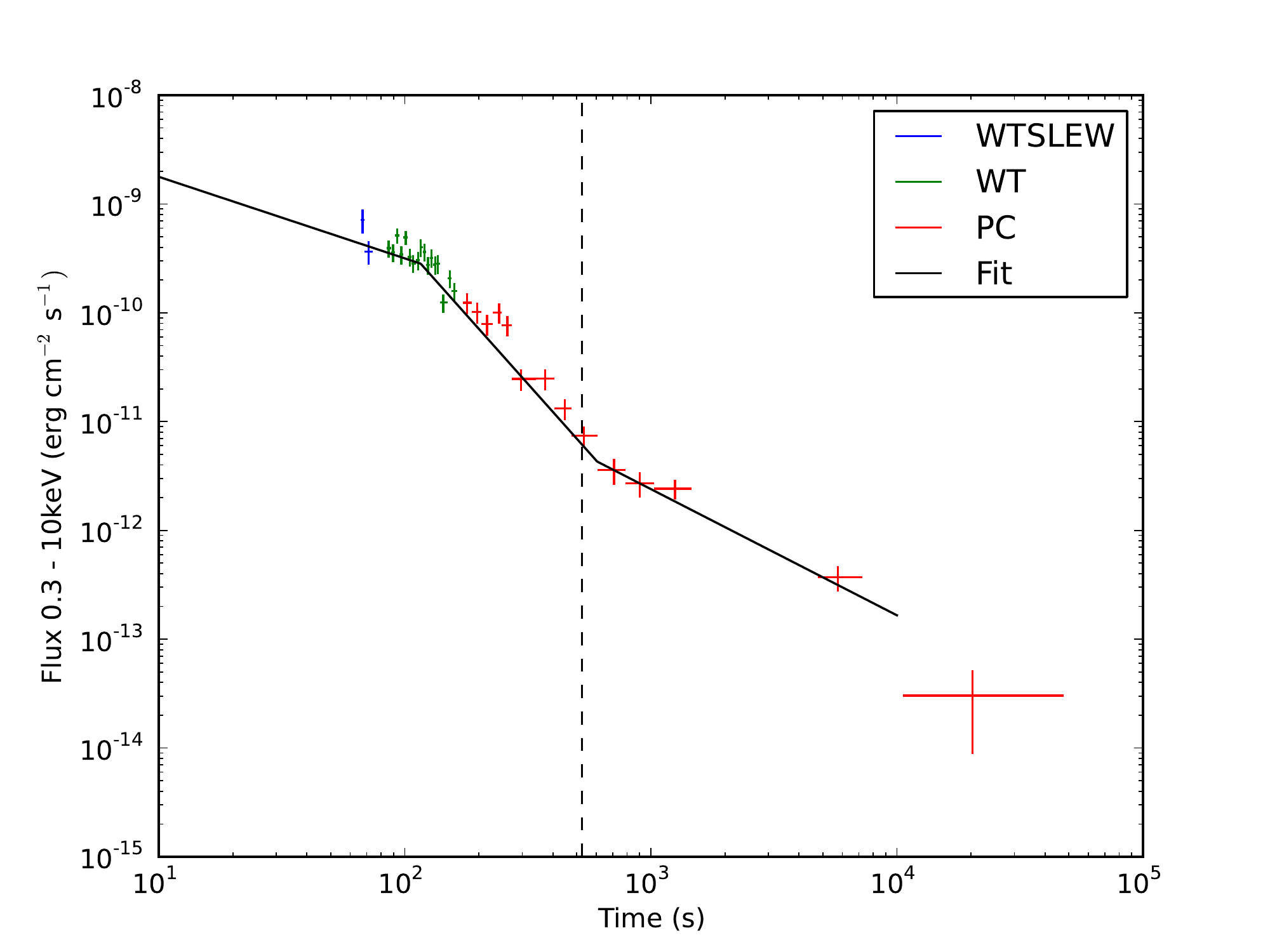}
\includegraphics[width=\linewidth]{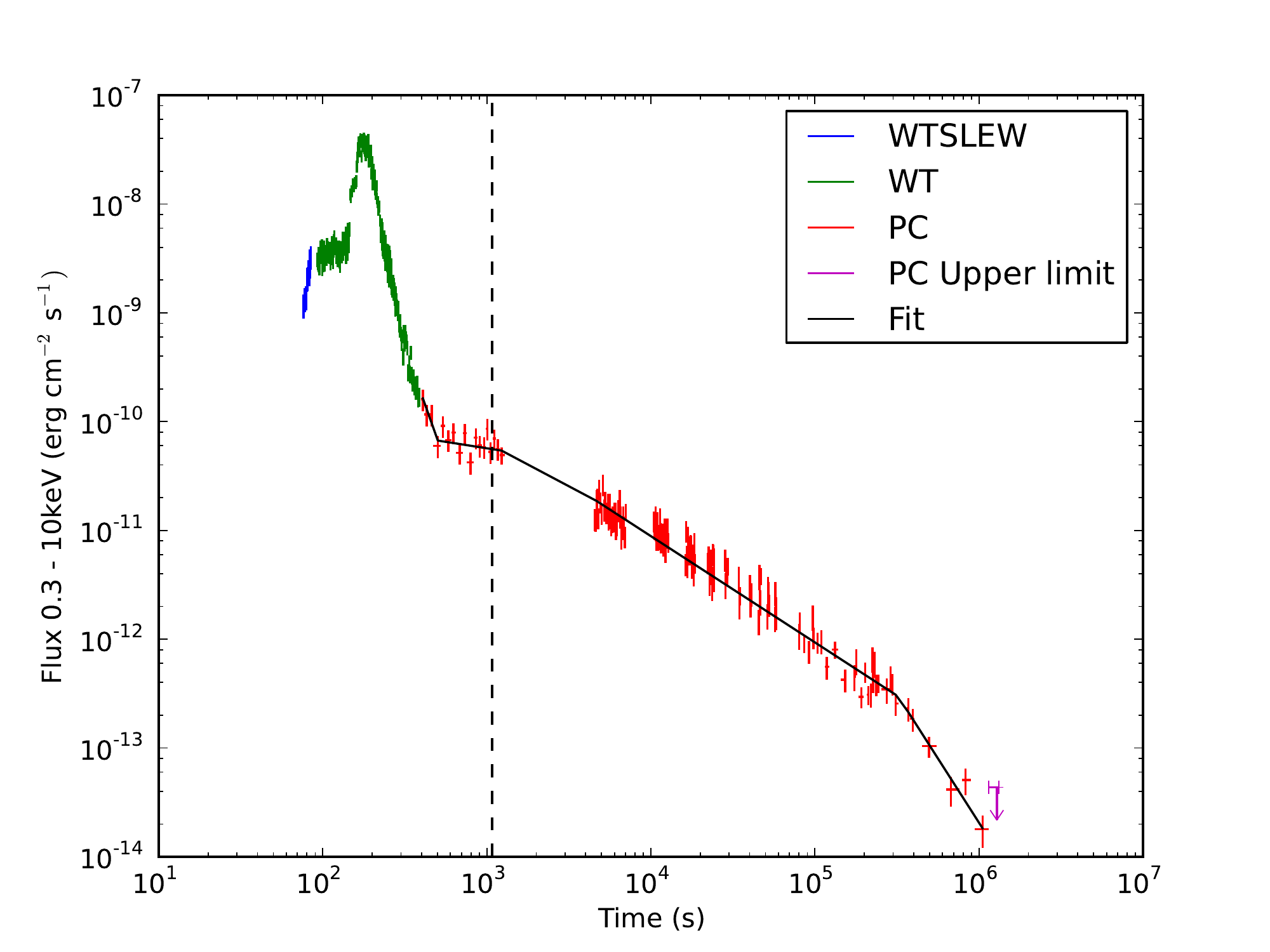}
\caption{The XRT (0.3$-$10~keV) light curves of the two GRBs after which we detected a single radio pulse. The arrival time of the radio pulse ($T_{\rm pulse}$; see Table \ref{tab:events}) is marked with a vertical line. Top panel: the X-ray light curve of GRB 101011A. The black line is fit is the fit derived by \citet{GCNR305}. Bottom panel: the X-ray light curve of GRB 10704A. The black line is the fit we derived. WTSLEW is the windowed timing mode during the slew of the spacecraft, during which flux measurements can be less reliable. WT is windowed timing mode. PC is photon counting mode.}
\label{fig:xrt}

\end{figure}

\emph{Swift/XRT} light curves are available for both GRBs for which we detected single pulses. For GRB 101011A, \citet{GCNR305} fit a three component broken power law to the 0.3-10~keV light curve, and derive a value of the final break in the power law of $t_{\rm break, 2} = 602^{+175}_{-88} \unit{s}$. \citet{GCNR305} do not describe their exact fitting method. We fit the same data with a broken power law, with each piece having a form $S(t) \propto t^{\alpha_i}$ for $0 \le i < N_{\rm break}$, and with each data point weighted by the flux error. We set the initial break times at $t=116$ and $t=602$, as derived by \citet{GCNR305}. From this fit ($\chi^2/N_{\rm d.o.f} = 1.6$) we derive a break time of $t_{\rm break, 2} = 707 \pm 173$. The arrival time of the radio pulse coincides with the break in the power law, to within the statistical errors for both fitting methods (Figure \ref{fig:xrt}). We note that the dispersion delay between the gamma-rays and our radio frequencies is less than 3~s and therefore insignificant for this calculation.

For GRB 100704A, no fit is derived in the GCN report \citep{GCNR293}. We fit a four component power law to the X-ray light curve for the data after $t=400 \unit{s}$ (Figure \ref{fig:xrt}), corresponding with the beginning of the proportional-counting mode data set. We used initial breaks times at 500,1000 and $3 \times 10^{5} \unit{s}$.   The resulting fit  ($\chi^2/N_{\rm d.o.f} = 1.2$), has a break at $t=1700 \pm 410$~s, which is within $1.5\sigma$ of the pulse arrival time at $t=1076 \unit{s}$. We note that the time of the fitted breaks is quite sensitive to the choice of initial break times.

\subsection{Periodicity search}
\label{sec:pulsar_search}

Our  method detected candidate repeating signals for all GRBs we observed, however, visual inspection of the candidate plots indicated that almost all of these candidates were associated with RFI.

The pulsar search yielded a single candidate for a repeating signal that was not conclusively associated with RFI. The candidate was detected in the data following GRB 110412A  with a S/N of 5.3 in the periodicity search, a topocentric period 524.48~ms and a DM of 51 pc cm$^{-3}$. After optimization for DM and period, the S/N of the candidate increases to 6.1. The candidate is brightest during the first 20 minutes of the observation. The signal is clearly detected in all sub-bands, as expected for a broad-band pulsar signal, and which also rules out narrow-band RFI. The profile comprises a single component with a width of about 5~ms . The signal is not very well defined in the period-DM plane or in the DM-S/N plot, but this is not unusual as the S/N is relatively low.

This position has been surveyed by the Parkes 70~cm survey \citep{Manchester96} with no detection. Given our sensitivity, any pulsar detected by our experiment would have been detected in that survey with very high significance, unless the pulsar was nulling at the time \citep{Backer70}, or the detection was a transient signal related to the GRB onset. It is common for low S/N candidates which appear convincing the first detection to not subsequently be detected in follow-up (M. Keith, private communication). On these grounds, the astronomical nature of this detection is very difficult to confirm and we will therefore not consider this candidate any further.

\subsection{Low-time resolution search}
\label{sec:long}

\begin{figure*}
\centering
\includegraphics[width=0.32\textwidth]{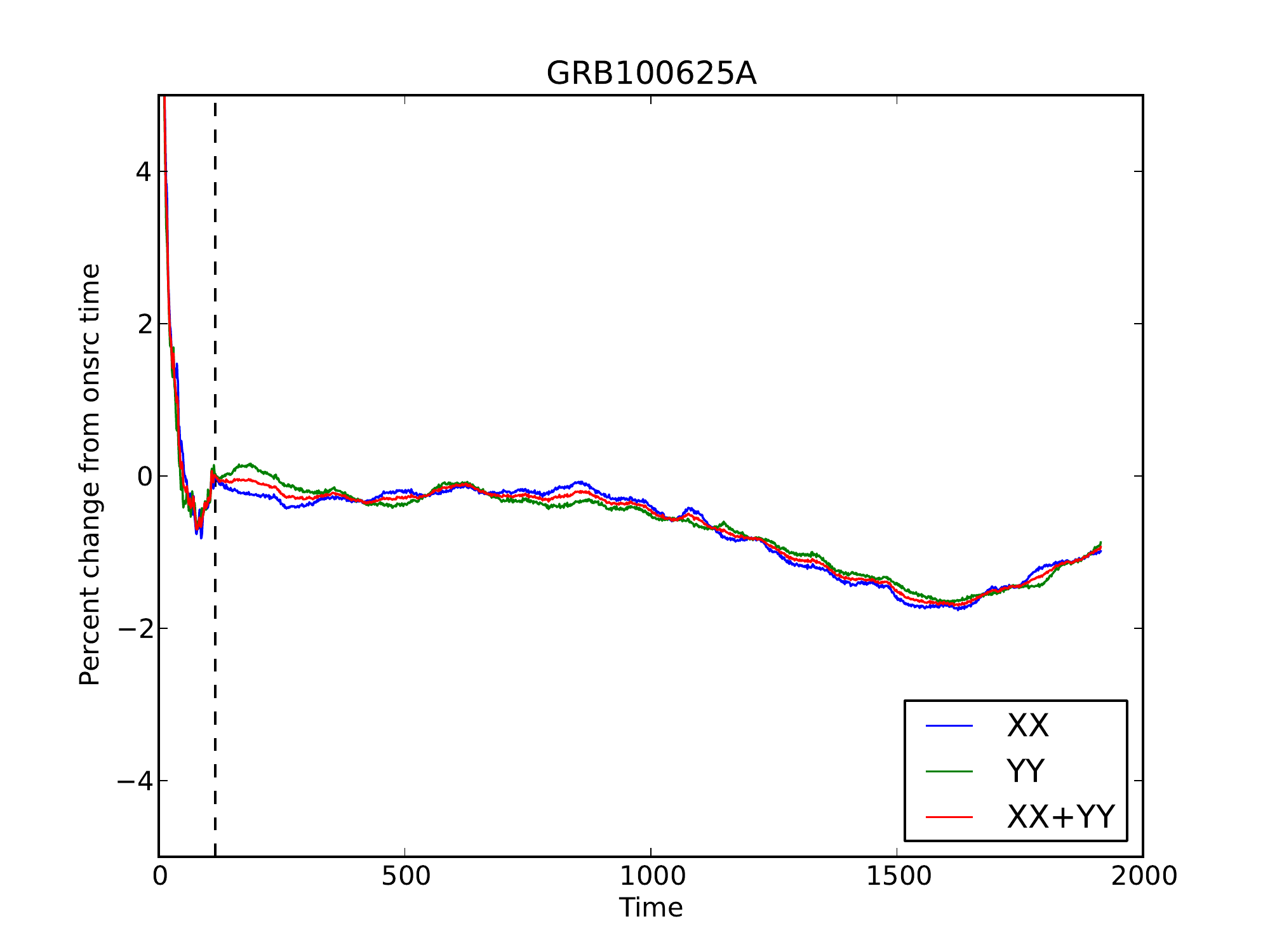}
\includegraphics[width=0.32\textwidth]{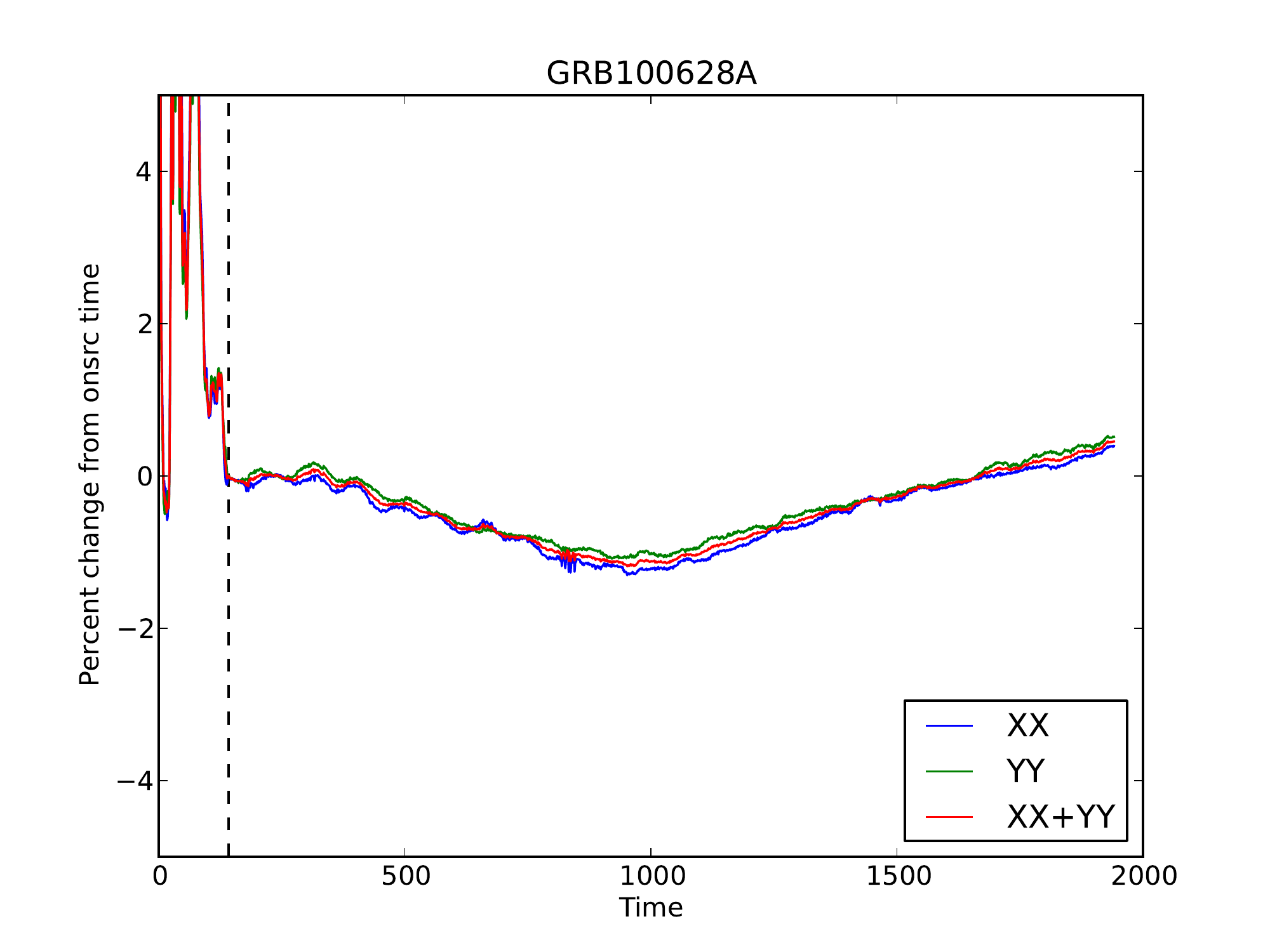}
\includegraphics[width=0.32\textwidth]{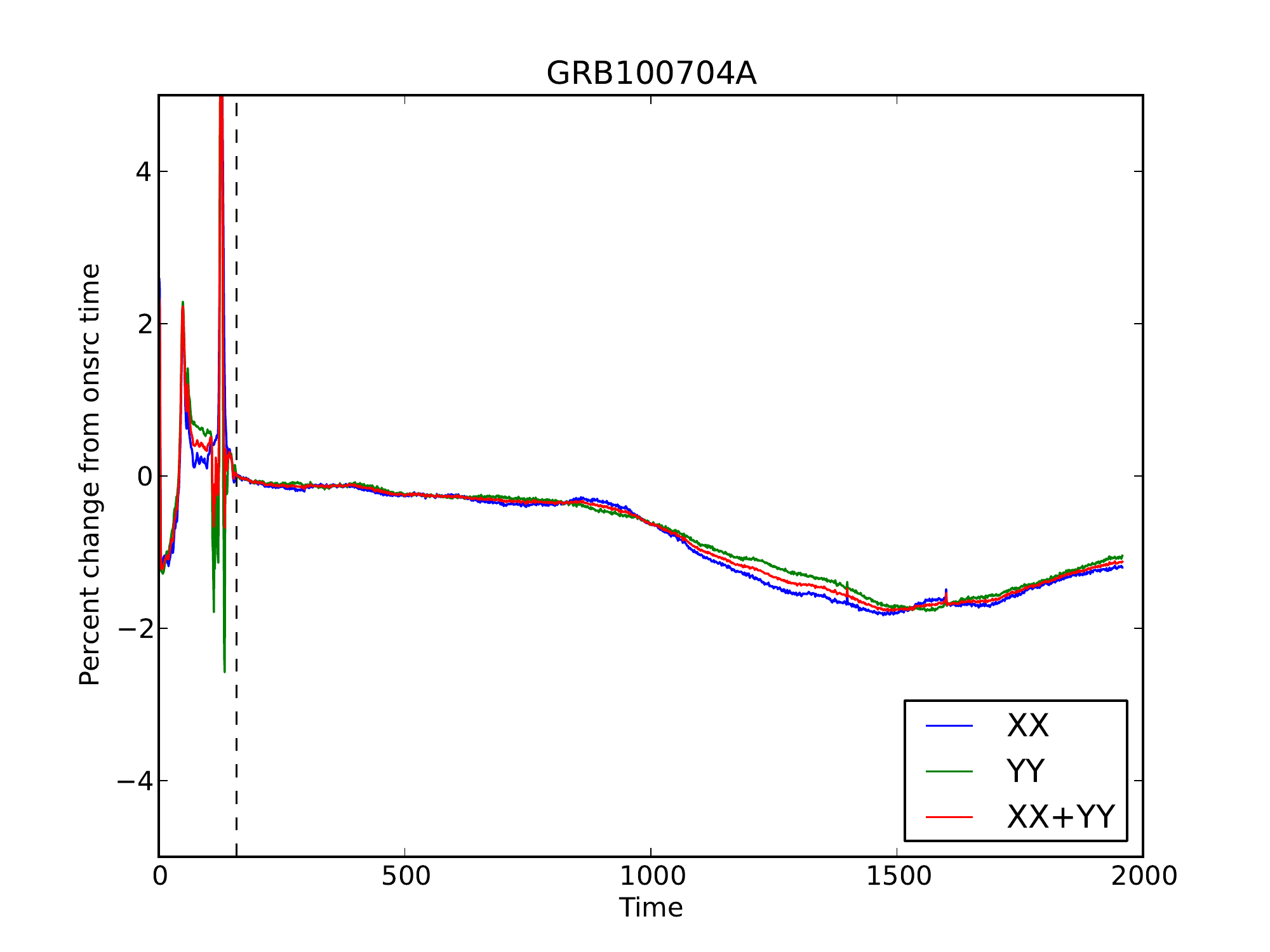}
\includegraphics[width=0.32\textwidth]{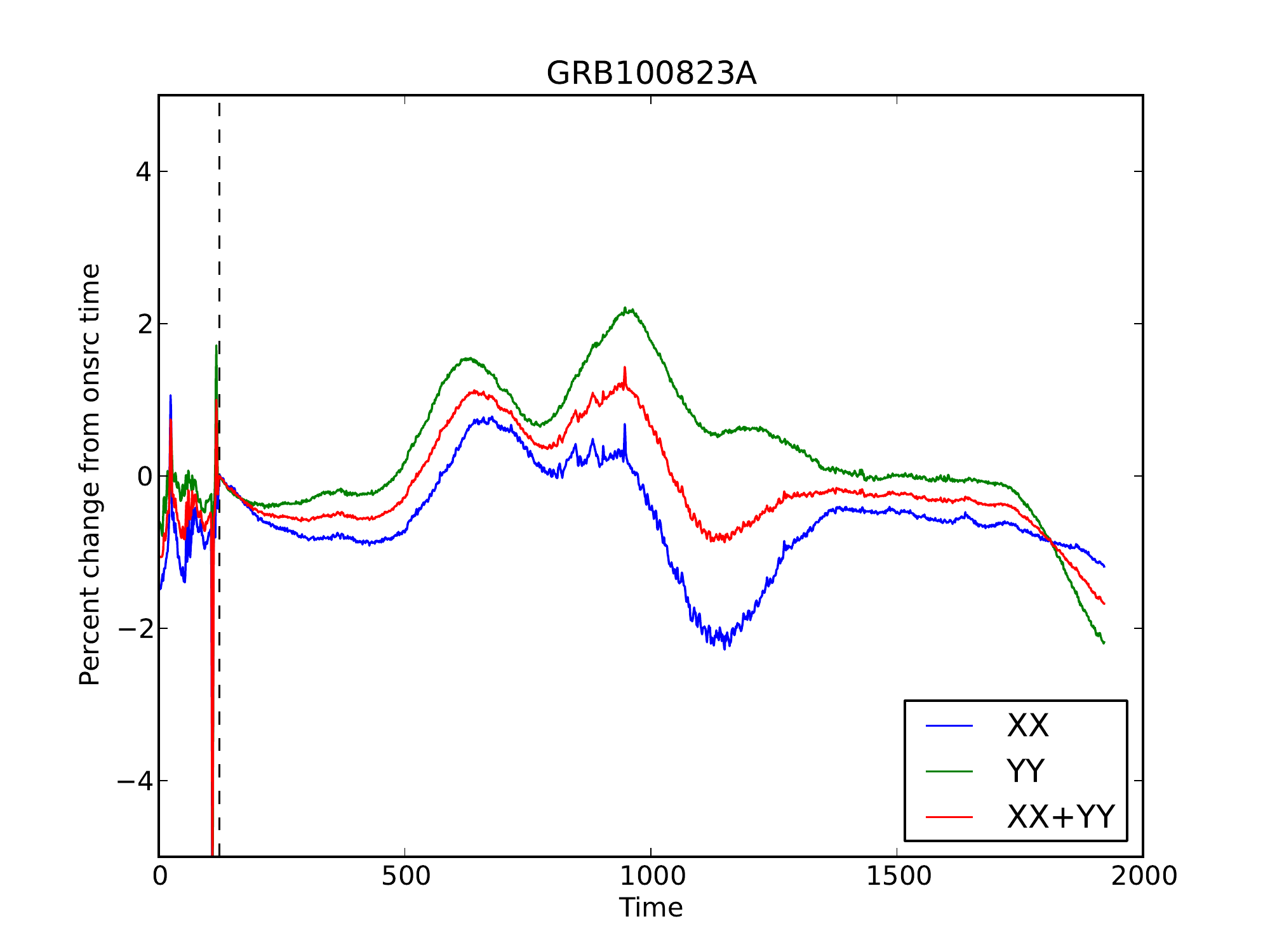}
\includegraphics[width=0.32\textwidth]{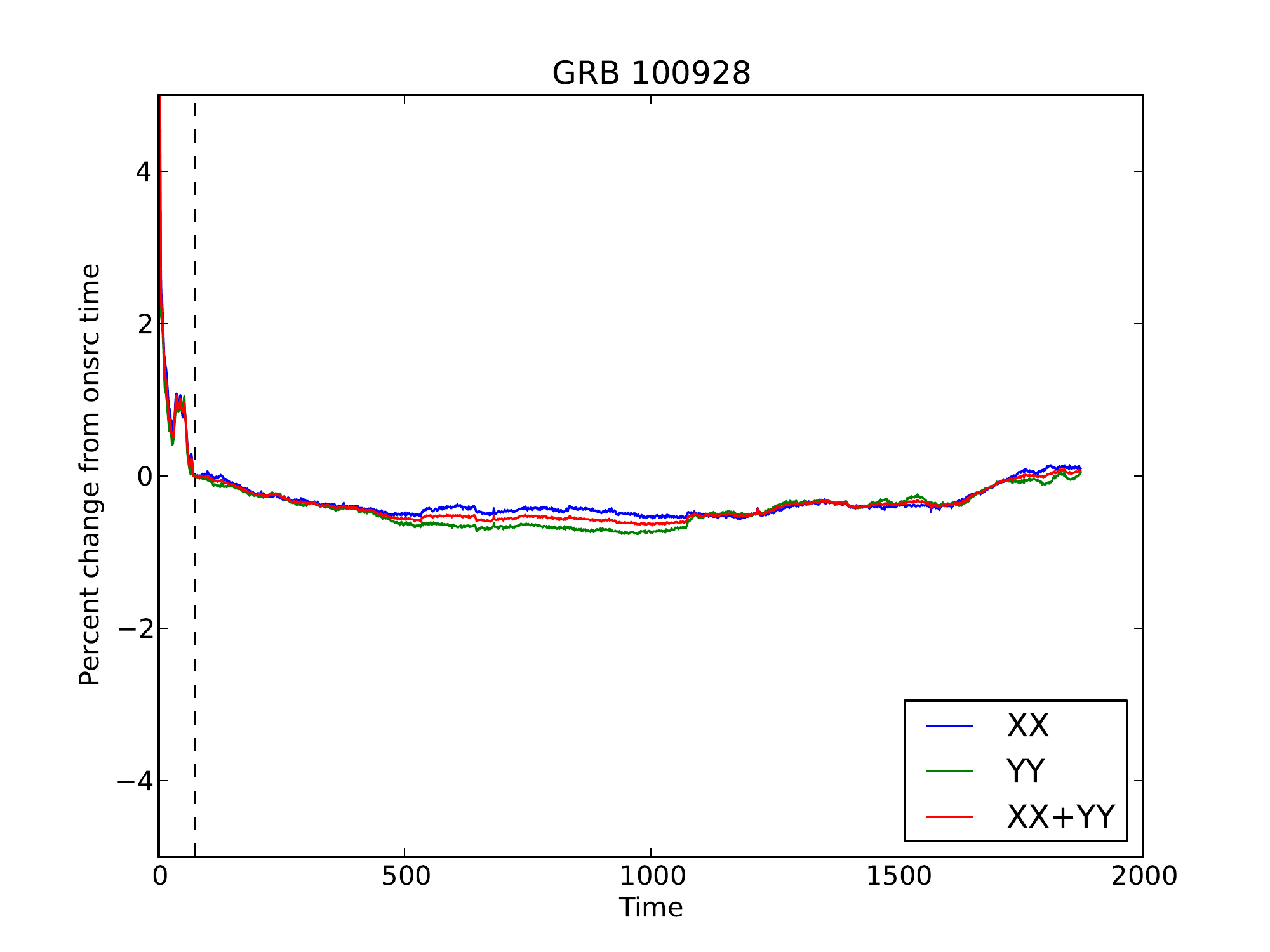}
\includegraphics[width=0.32\textwidth]{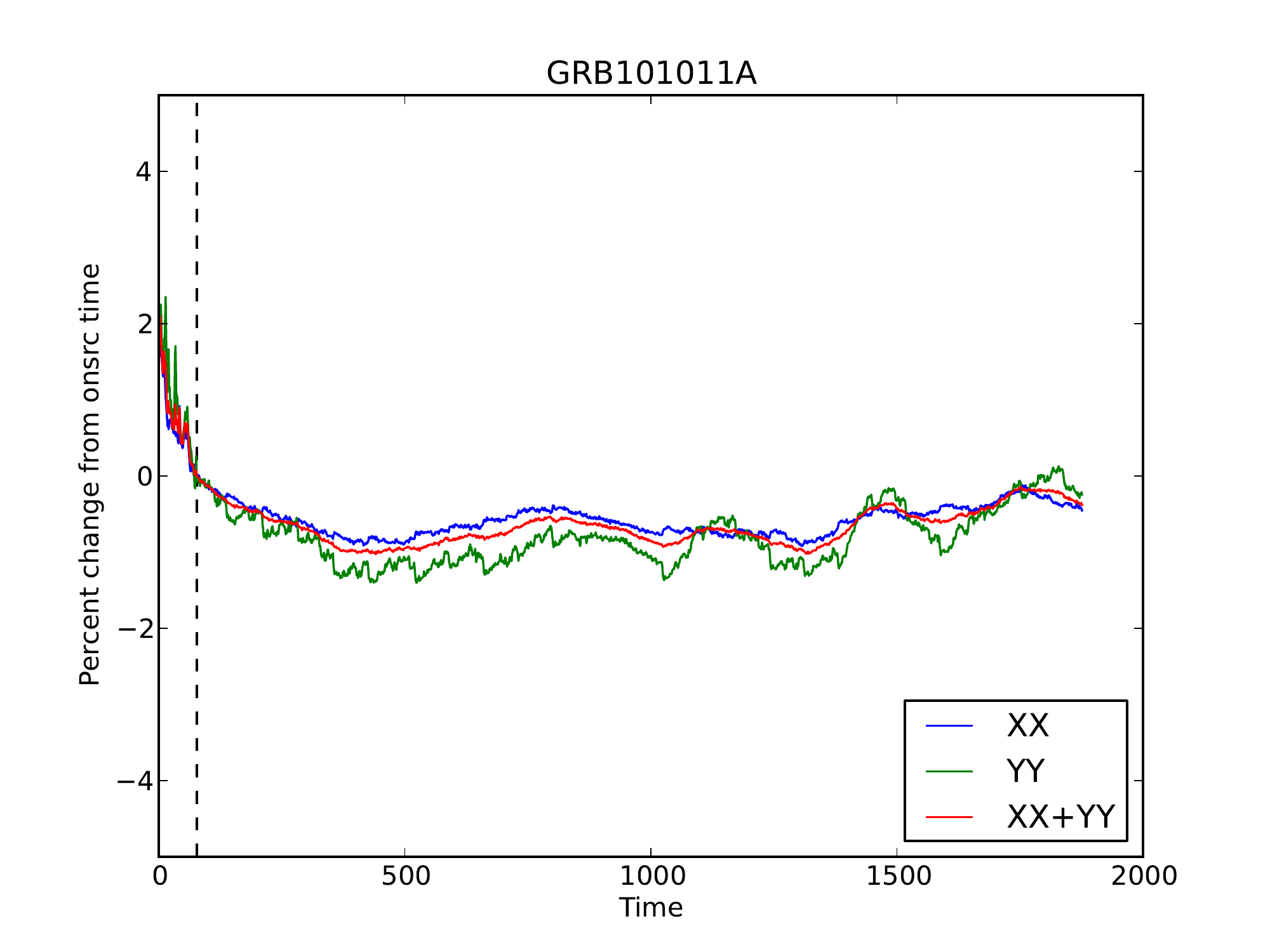}
\includegraphics[width=0.32\textwidth]{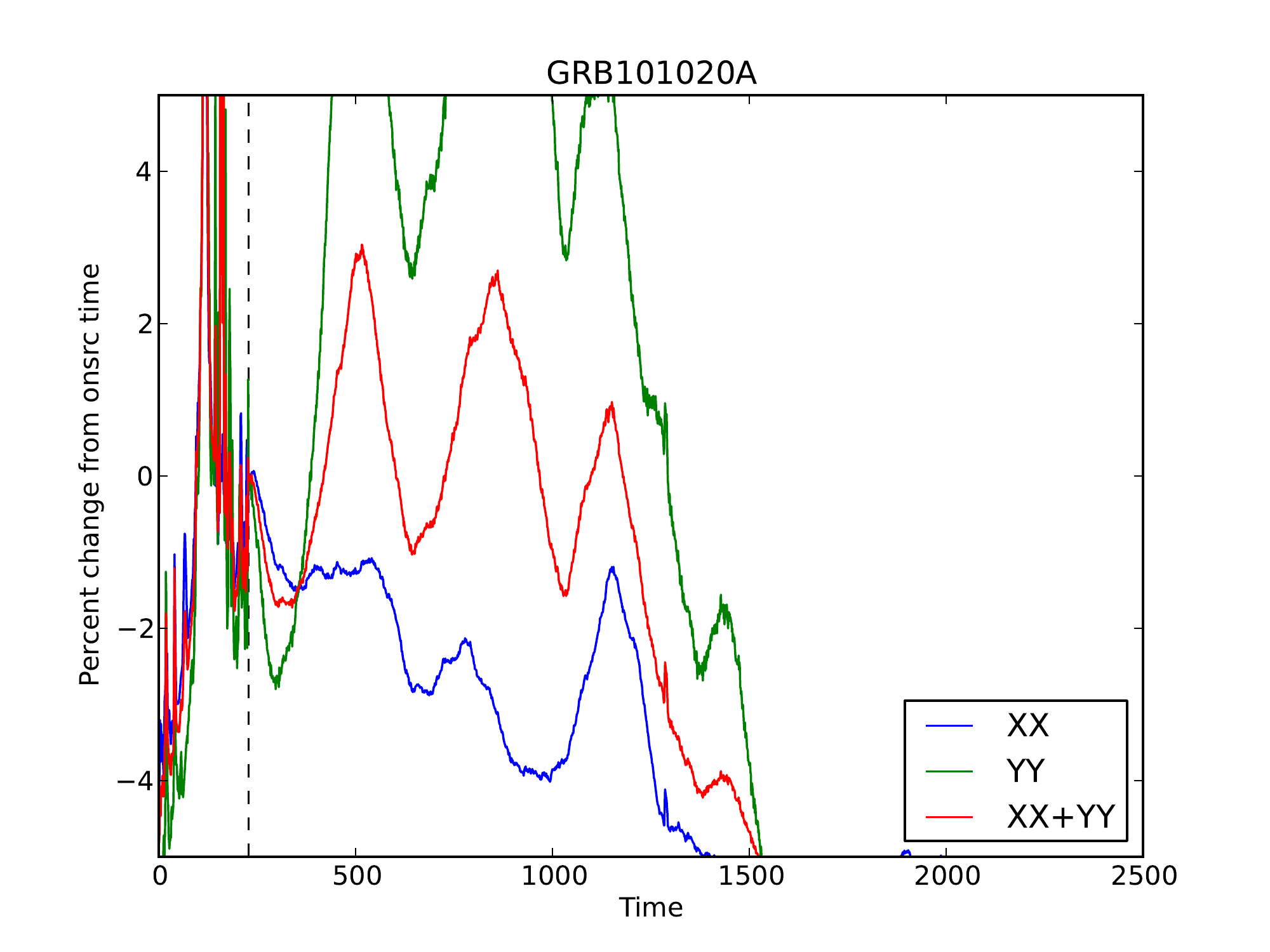}
\includegraphics[width=0.32\textwidth]{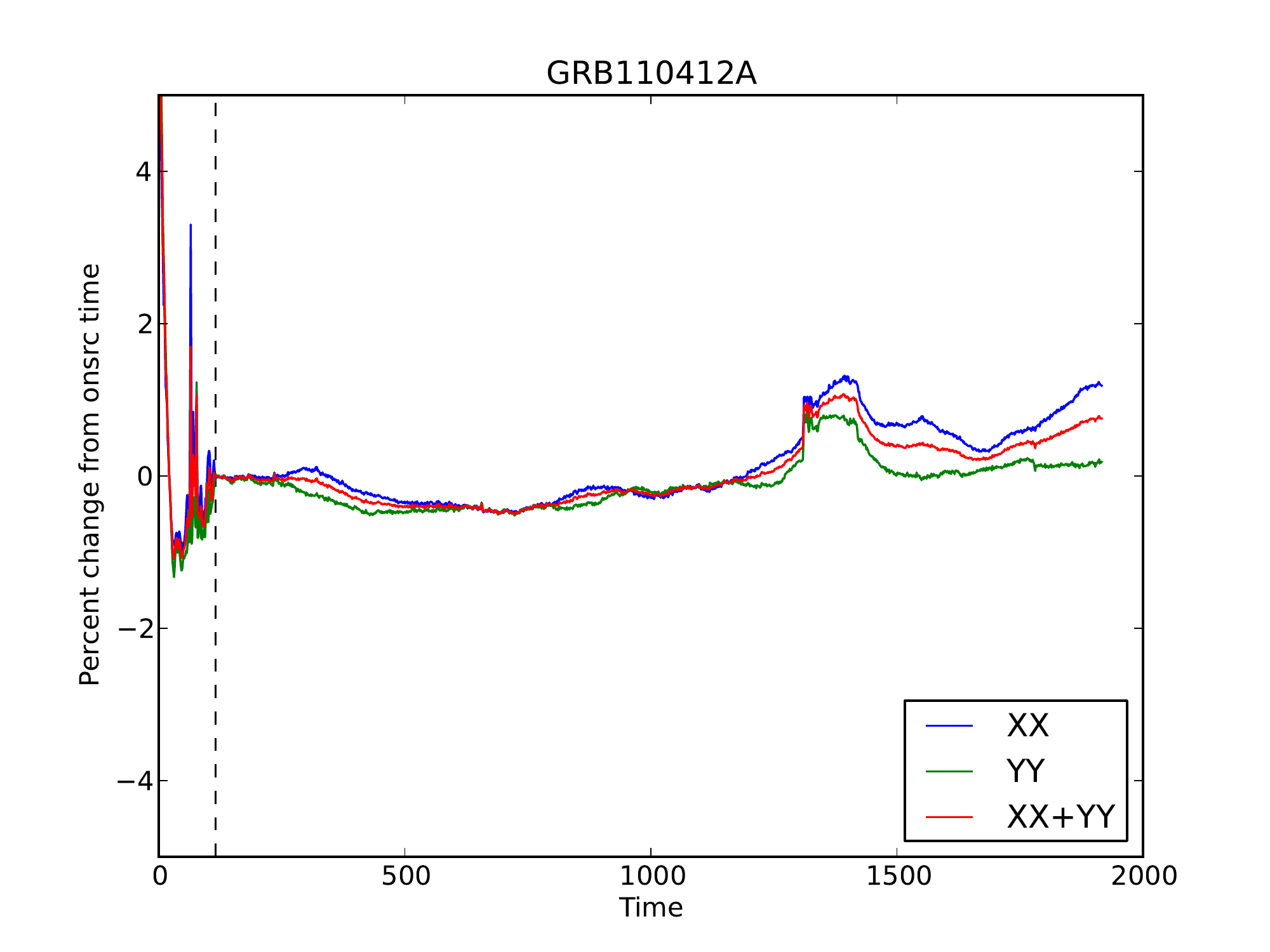}
\includegraphics[width=0.32\textwidth]{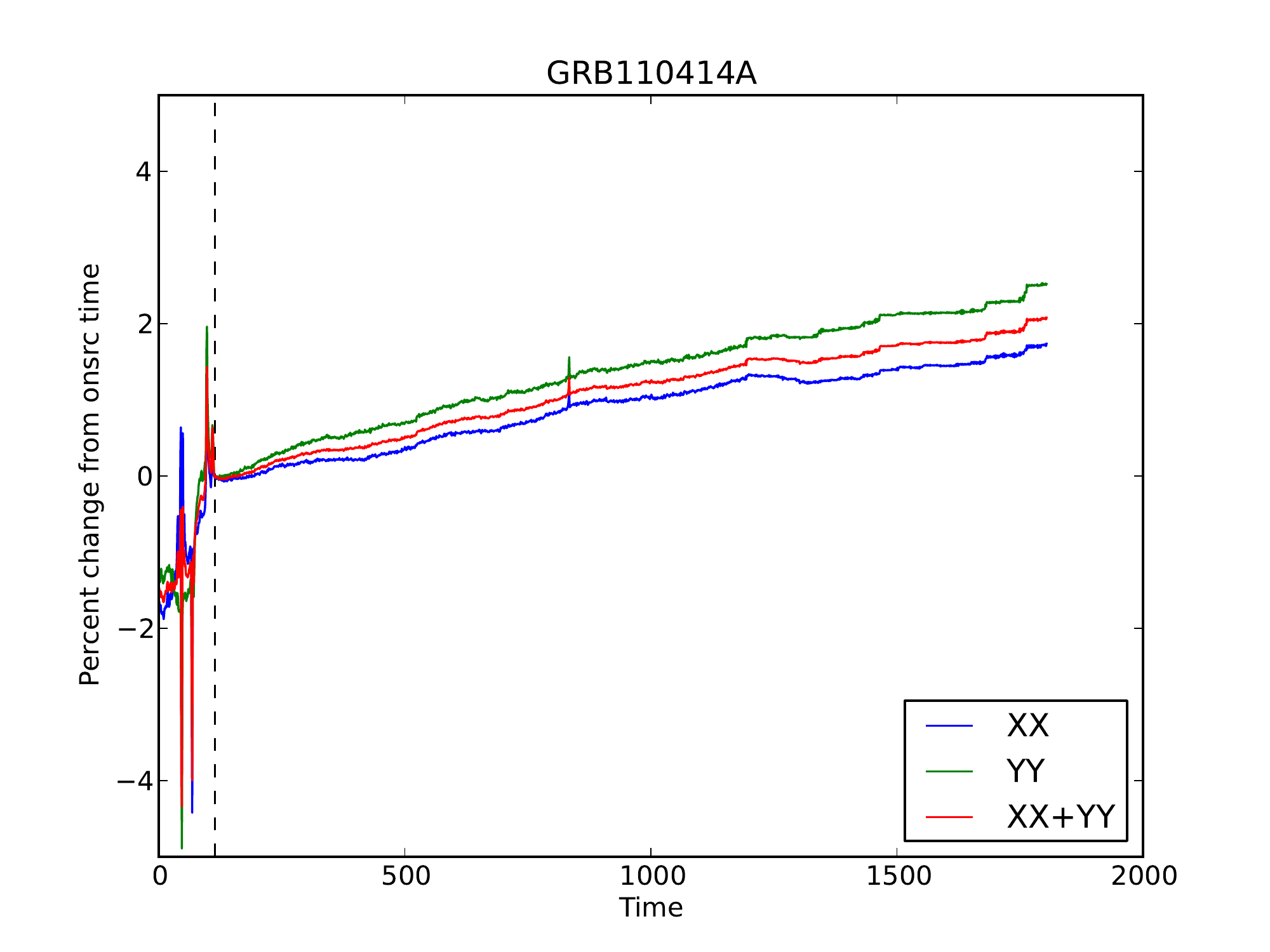}
\caption{Low-time resolution 1.4~GHz light curves for all GRBs. The vertical dashed line indicates when the antenna was on source, with the interval preceding this line being measurements during the slew.}
\label{fig:relpwr}
\end{figure*}

\begin{figure}
\centering
\includegraphics[width=\linewidth]{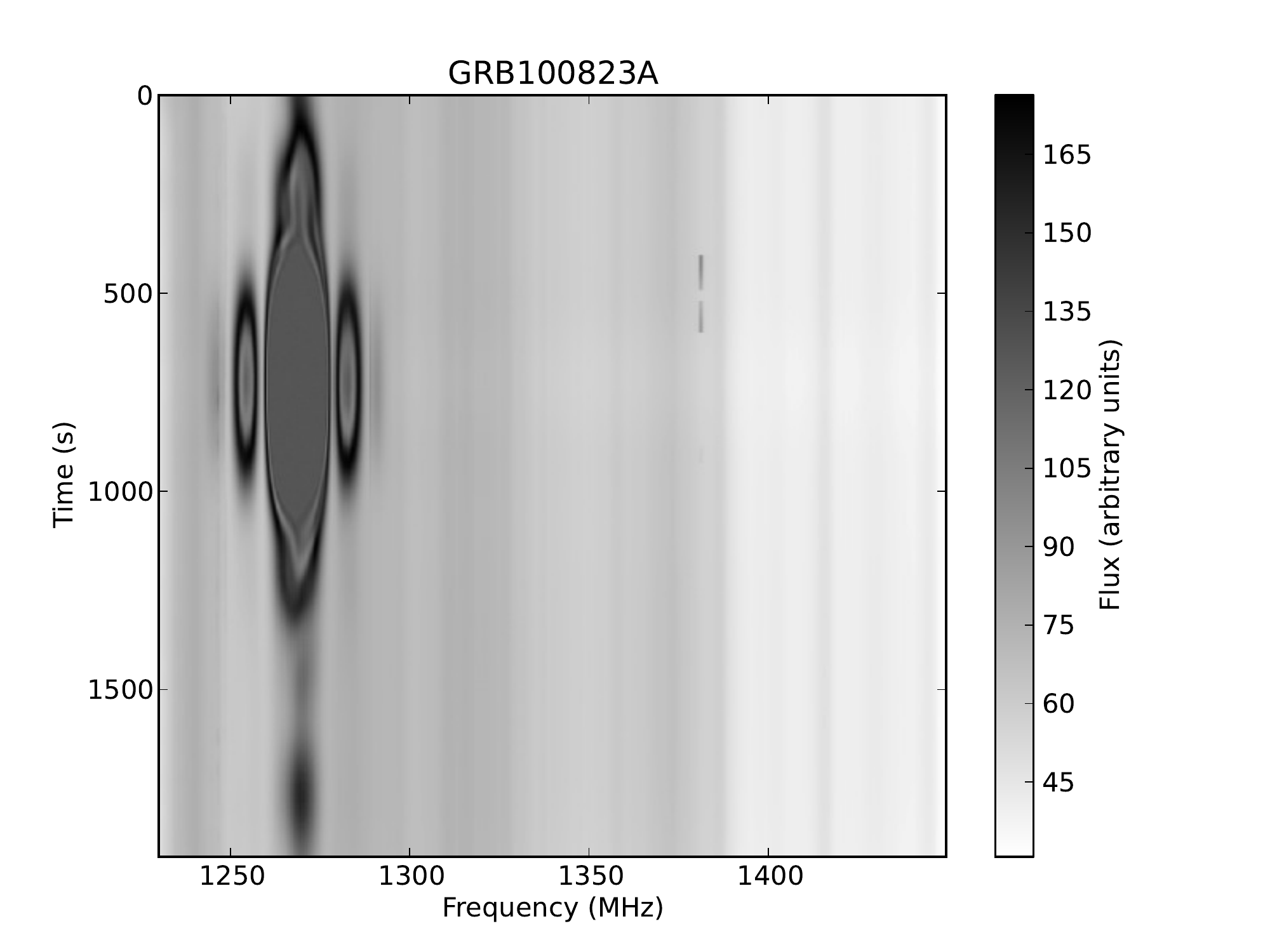}
\caption{The raw receiver bandpass during observations of GRB 100823A, as a function of time and frequency. The gray scale is an arbitrary linear scale of the intensity. Long time integration measurements  were badly affected by RFI from the \emph{Beidou G1} satellite, despite zero weights being applied to the obvious interference.  The interference was centered on 1270~MHz, the transmitting frequency of the satellite, and the interference was worst when the pointing direction passed closest to the satellite.}
\label{fig:432420_pols}
\end{figure}

The light curves at 1~s integration are shown in Figure \ref{fig:relpwr}. The light curves with rapid changes in flux density can be attributed to interference. For example, the flux density variations during observations of GRB 1000823A are due to RFI from the \emph{Beidou G1} satellite.\footnote{Refereed, English information about the \emph{Beidou} global satellite system is non-existent. The only available information available in English is on general news and satellite navigation websites. Perhaps the best source of information can found at \url{http://en.wikipedia.org/wiki/Beidou_navigation_system}. The official Chinese website is \url{http://www.beidou.gov.cn}. \emph{Beidou} satellite ephemerides are provided by the US Space Command and the real time position was computed using data provided at \url{http://www.n2yo.com/}.} This can be seen from the modulation spectrum centered at 1270~MHz  in the bandpass and variance plots. During the observations, the telescope pointing direction was within 10 degrees of the satellite position when the interference was most severe (Figure \ref{fig:432420_pols}). Although our method flagged the worst channels, it is clear that additional power leaked through the unflagged channels, affecting the light curve.

The substantial flux variations during observations of GRB101020A can be attributed to the proximity of the Sun to the observing direction, which was 25 degrees from the Sun at the beginning of the observation. None of the narrow-band RFI coincides with the large changes in flux density, indicating that the changes in flux are broad-band and are likely to be a result of the Sun moving through the far sidelobes.

\begin{figure}
\centering
\includegraphics[width=\linewidth]{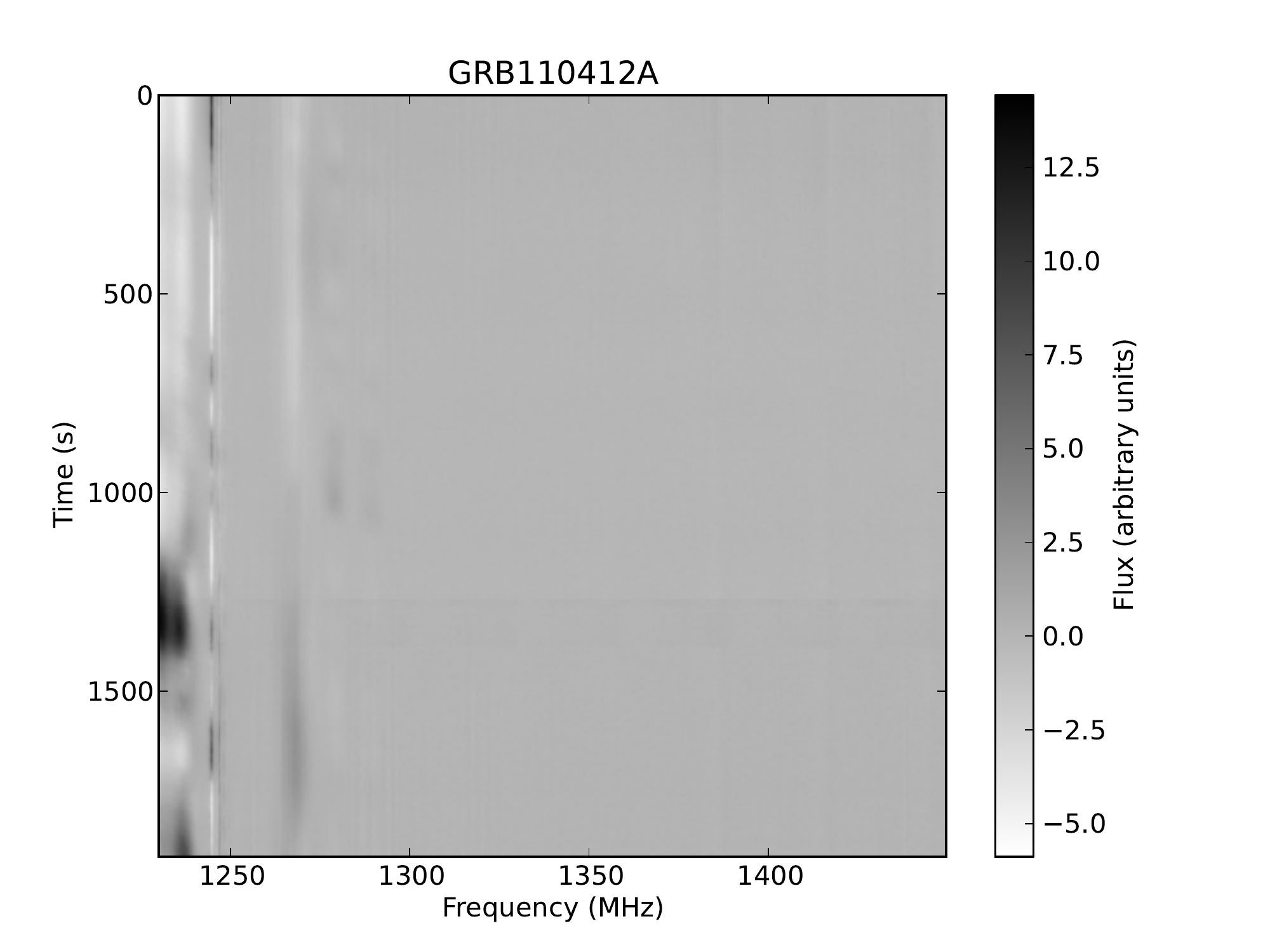}
\caption{The baseline-subtracted receiver bandpass during observations of GRB 110412A, as a function of time and frequency. Long time integration measurements of  were affected by narrow band RFI at around 1230~MHz. The narrow-band interference was worst at around $t=1300 \unit{s}$, which corresponds to a bump in the broadband lightcurve.}
\label{fig:451191_pols_biased}
\end{figure}

The variations during observations of GRB 110412A are likely due to unflagged narrow-band interference at around 1230~MHz (Figure \ref{fig:451191_pols_biased}). These channels had no substantial kurtosis and so were not flagged, but had slightly higher variance than the rest of the band (Figure \ref{fig:rfi}). Nonetheless, the fact that the interference is narrow band strongly suggests RFI as the source of the variation.

The slow variations (e.g. GRB 100625A) are due to gain variations in the system, which is stable to about 5\% in 30 minutes. We were unable to calibrate out these gain variations, as our feed was not equipped with a switched radiometer.

\section{Discussion}

\subsection{Single-pulse detections}

We have two single-pulse candidates, the brightest of which is a $6.6 \sigma$ pulse 524~s after GRB 101011A. Below we discuss possible origins for this pulse. Our  conclusions also broadly apply to the other candidate single-pulse (the $6.2 \sigma$ pulse 1076~s after GRB 100704A), but we restrict our attention to the brightest pulse.

The DMs of both pulses are much larger than the DM in the GRB direction predicted by the NE2001 model \citep{Cordes02} for our Galaxy. For GRB 101011A, NE2001 predicts a maximum DM of 39~\pccm at a DM distance of 3.14~kpc, which implies an excess DM of 530~\pccm. For GRB 101011A, NE2001 predicts a maximum DM of 106~\pccm at a DM distance of 9~kpc, which implies an excess DM of 89~\pccm. The large excess DMs for both radio pulses are well above the uncertainties in the NE2001 model and therefore, if the pulses are indeed astronomical, exclude Galactic sources such as pulsars or RRATs. The excess DM must therefore be due either to intergalactic material or a possible host galaxy.

\subsubsection{Random noise fluctuation - theoretical arguments}
\label{sec:random_theoretical}

In principle we can apply statistical arguments to derive the probability of a $6.6\sigma$ pulse occurring by chance in our data. In practice, such arguments are not straightforward for a number of reasons. Firstly, our data are not easily described by analytic probability distribution functions, chiefly because of the presence of RFI. Secondly, the processing pipeline applies a number of RFI mitigation strategies and matching algorithms that are difficult to model in a statistical sense. Finally, the search over a range of DMs and pulse widths, which necessarily sample overlapping regions of parameter space, is not independent. That is, a genuine astronomical signal will appear in a number of adjacent DMs and boxcar trials, implying that the simplifying assumption of independent trials is not valid.

Nonetheless, we can construct a simple model with which to evaluate our real-world results, assuming Gaussian noise and independent trials. Here we follow the logic of \citet{BurkeSpolaor11}. For a 30 minute observation, we produce 1991 DM trials, each comprising $N_s =28.125 \times 10^6$ samples. We apply a set of 9 boxcar trials to each DM trial, implying that a total of $N_p =1991 \times \sum_{i=0}^{9}{N_s/2^i} \simeq  1.1 \times 10^{11}$ points are searched for pulses. Assuming each trial is independent, then the expected number of false alarms exceeding a threshold $T$ during an observation is $ N_p \textrm{erfc} (T/(\sigma \sqrt{2}))$, where erfc is the complementary error function.

Applying this model to observations of GRB 101011A, we expect $N_{6 \sigma} \simeq 216$ trials above $6 \sigma$ and get 260 such detections, roughly tallying with our discussion to this point, but many detections are clearly associated with RFI and are not related to the Gaussian distribution. At the $6.6 \sigma$, significance of our detected pulse, we expect on average 4.6 false alarms per observation. Applying the Poisson distribution, we can compute the probability of finding exactly one false alarm during a 30 minute observation, at or above $6.6 \sigma$, of 4.6\%. Alternatively, we can compute the probability of finding one or more pulses above $6.6 \sigma$ of 94\%.

These relatively high probabilities suggest that it is quite likely that we would find a $6.6\sigma$ candidate in a single observation, and that we would also expect several candidates across our 9 observations, exactly matching our results. However, we have neglected up to this point the `friend-of-friends' algorithm employed by the single-pulse pipeline, which declares a candidate as viable only if a pulse is detected in 3 or more nearby DM trials. According to \citet{BurkeSpolaor11}, this algorithm reduces the number of candidates by a factor of $\simeq 100-1000$, implying that applying the algorithm reduces the probability of detecting exactly one candidate from 4.6\% to 4.4\%. Such an analysis suggests that the $6.6\sigma$ pulse we have detected is not a random noise fluctuation, with a confidence of greater than 95\%.

Relaxing the assumption of independent trials implies that adjacent trials (either in DM of boxcar width) have overlapping time and frequency samples. Thus, if a given trial has a high value, an adjacent trial is also likely to have a high value, because many of the samples overlap. Therefore, relaxing the assumption of independence increases the false alarm probability, and we would expect to see clusters of false alarms in DM and boxcar space. The amount of overlapping samples between adjacent trials is of the order of 50\% both in time and frequency. For a trial with a $6\sigma$ detection, which is already  unlikely, the probability of the non-overlapping samples from an adjacent trial pushing the adjacent trial above $6\sigma$ is small. As the friend-of-friends algorithm requires three or more adjacent trials  above $6 \sigma$ to declare a detection, it effectively nullifies the correlation between adjacent trials, and the probabilities described above are still accurate.

For simplicity, we have neglected the effect of the additional weak RFI filter, which discards candidates with adjacent low DM pulses within 3~s (Section \ref{sec:htr_processing}). Such a filter will reduce the number of candidates produced by random fluctuations even more, which in turn increases the confidence.

\subsubsection{Random noise fluctuation - A null trial}

The theoretical considerations in Section \ref{sec:random_theoretical} assume Gaussian noise, and do not deal with the complexities of the RFI and detection algorithms very adequately. Perhaps a more reliable approach to determining the probability of such a pulse occurring by chance is to measure the number of false positives recorded when there is no GRB in the beam. One such method would be to observe a `blank' patch of sky, not containing a GRB, for a period of time, and measure the number of candidates. Unfortunately, such observations were not made when the antenna and feed were available.

Ideally, we would like to remove the astronomical signals from the data, while preserving the noise properties and RFI, so that an accurate measurement of the background can be performed. Once such approach is to search for pulses in the data we already have, but randomize the channel ordering before dedispersion. Randomizing the channels destroys the $t \propto \nu^{-2}$ dispersion relationship expected from an astronomical pulse, and almost guarantees that no genuinely astronomically dispersed pulses will be visible in the results, especially at large DMs. Randomizing the channels still preserves the noise properties, and narrow-band, and broad-band impulsive RFI, as desired, but impulsive RFI with intermediate bandwidth is destroyed. The level of intermediate bandwidth RFI is therefore reduced in the randomized data, and the number of background detections is consequently reduced. We expect the amount of intermediate bandwidth RFI to be small in comparison with the broad-band RFI, so we expect this effect to be small.

This technique, while inferior to taking additional data on a blank patch of sky, still has advantages over the theoretical approach. In particular, it uses the same detection algorithms as those used on the on-sky data, the noise statistics of the data are preserved, and the zero DM sequence (which is used for RFI flagging by the pipeline) is also preserved.

We conducted such a trial based on the data from GRB 101011A (which contained the brightest single pulse). A software-based pseudo-random number generator was seeded with  random data from the Linux kernel entropy pool at the beginning of each observation, and the ordering of the channels was changed randomly every 25~s during the observation.

We simulated 248 30 minute observations with the randomized channel ordering, which yielded seven single-pulse candidates. From this trial, we can conclude that the probability of detecting a $>6 \sigma$ candidate that passes all the additional criteria is approximately 3\%, and we therefore rule out the possibility of our pulse being due to random fluctuations with a confidence of 97\%.

\subsubsection{Impulsive RFI}
\label{sec:pulse_rfi}

Impulsive RFI has a number of attributes that make it often distinguishable from astronomical sources: it is usually most visible at low DMs, although it can extend to higher ones, depending on the exact waveform shape. Impulsive RFI can often be clustered in time, with many RFI events occurring frequently over periods of minutes, often followed by similar periods of no RFI. For example, the observations of GRB 101011A have a single RFI event in the first 15 minutes observing, and 38 RFI events in the second 15 minutes (Fig. \ref{fig:search_plots}).

The pulse seen just after GRB 101011A has a very high DM of 570 pc cm$^{-3}$, which argues against an RFI origin. The nearest low DM detection (a $>6 \sigma$ isolated pulse) appears to be some 120~s before the candidate pulse, which argues against the candidate being related to a generally high level of RFI at the time, or being a high-DM component of RFI which is also detected at low DMs (Figure \ref{fig:search_plots}).

The high DM, absence of contemporaneous low-DM RFI add credence to a non-RFI origin. However, in the absence of a coincidence detection at a geographically distant site, we cannot definitively rule out RFI as the origin.

\subsubsection{Lorimer burst events}

Our  pulse has similar properties to the Lorimer burst (LB) of \citet{lorimer2007bmr}. The DM of our pulse is higher than the LB,  but this can be explained by differences in Galactic electron content, redshift and ionized material near the source. The Lorimer burst is shorter than our pulse (5~ms for the LB as opposed to 25~ms for our pulse). This shorter duration cannot be explained by reduced scatter broadening from a lower electron column, implying that the difference in pulse duration is  intrinsic.  The flux densities are roughly comparable (30~Jy for the LB, 7~Jy for ours). If the LB is associated with a GRB then it is likely that we have observed the same physical mechanism.

\subsubsection{``Perytons''}
\label{sec:perytons}

In an archival search of Parkes Multibeam archive, \citet{BurkeSpolaor11lg}  discovered 16 single pulses with similar spectral characteristics to those expected from a  dispersed extragalactic pulse. These pulses, that \citet{BurkeSpolaor11lg} call `perytons', have a number of unique properties. In general, the perytons follow a $t \propto \nu^{-2}$ frequency sweep, although some pulses have significant departures from this sweep. The sweep times across the 288~MHz band centered at 1374~MHz are between 200 and 400~ms, but mostly clustered in the range 320-400~ms. The perytons are extremely bright (between $0.1-272$~kJy), and have substantial amplitude modulations. All perytons were detected during the day, in general mid-morning, and mostly in the mid-winter months of June and July. The peryton event rate is $2.3 \times 10^{-7} \unit{deg^{-2} hr^{-1}}$. \citet{Kocz12} find that the perytons are most likely local RFI, as they tend to arrive in the second half of a human second, and arrive with a characteristic time gap of $\sim 22$~s.

The pulses we detect have somewhat different properties. Unfortunately, we do not have sufficient S/N to detect small departures from the cold plasma dispersion law, nor any substantial amplitude modulations. Figure \ref{fig:436094_pulse}  supports a standard cold plasma frequency sweep, and constant amplitude, although the significance is not very high. The flux densities of our pulses are approximately 7~Jy, substantially fainter than the perytons. If we extrapolate the sweep time of our brightest burst for the configuration of \citet{BurkeSpolaor11lg} between 1230 and 1518~MHz, we obtain a sweep time of 536~ms. This sweep time is substantially higher than that of any of the known perytons, but within the general range, given the small number statistics. Our  brightest pulse was detected on 2011 Oct 11, well outside the usual Peryton season, and at 22:58 (local time) in the evening, outside the usual Peryton time. There had been no rainfall and no lightning detected within 50~km of the telescope within 24 hrs of the GRB time, disfavoring the postulated lightning related cause.  Finally, the event rates are vastly different, with our events being more common by over six orders of magnitude, although \citet{BurkeSpolaor11lg} note that the Peryton event rate is highly non-uniform, with 12 perytons detected within 0.5 hr in one instance.

From this analysis we conclude that the substantially higher event rate, longer sweep time, arrival date, time of day, and prevailing weather conditions of our pulse all make it unlikely to be a peryton.

\subsubsection{Gamma-ray bursts}
\label{sec:grb}
Two out of eight\footnote{The observation of GRB 101020A  is not considered in these statistics, due to RFI.} (25\%) of our observations contain a 1.4~GHz radio pulse, which begs the question, are the GRBs themselves the source of these pulses?  Unfortunately, there is no known physical mechanism for producing such a burst within minutes of a GRB, in spite of the substantial theoretical work inspired by the detection of the LB.

Any proposed mechanism must explain a range of properties. First, our two detections were only present following long gamma-ray bursts, so a mechanism must fit within the confines of the collapsar paradigm \citep{MacFadyen99} (see below). The pulses are detected with  flux densities of the order of 7~Jy, which requires a very powerful mechanism if placed at the cosmological distances of GRBs.  The short pulse width, of the order of 25~ms, requires a relatively compact causative region. Our pulses are 524 and 1076 seconds after the gamma-ray triggers, much later than the $\sim 2$~s dispersion delay, so any mechanism must explain this delay.  Finally, any mechanism must also explain why both pulses are coincident with a break in the X-ray light curve, and also must explain the complete lack of flaring at X-ray or gamma-ray wavelengths simultaneous with the emission of the radio pulse.

Unfortunately, no host galaxies were observed for either GRB, so no redshift is available. For the remaining analysis, we assume a redshift of $z=2.8$, which is approximately the mean redshift of GRBs from \emph{Swift} \citep{Jakobsson06}.

We now try to interpret our pulses within the context of the collapsar model \citep{MacFadyen99}, namely: the collapse of a massive ($> 10 \unit{M_{\odot}}$) star producing a central engine, either black hole, or magnetar \citep{Usov92}, of mass $\sim 2-3 \unit{M_{\odot}}$; which in turn accretes material from a disk which produces jets, the prompt gamma-ray emission, and the shocks that pass through the surrounding material and into the interstellar medium.

Concentrating on the brightest of our radio pulses, the flux density of approximately 7~Jy implies a spectral luminosity of $1.1 \times 10^{36} \unit{erg~s^{-1}~Hz^{-1}}$ at redshift 2.8 ($H_0 = 71 \unit{km s^{-1} Mpc^{-1}}$, $\Omega_M = 0.27$, $\Omega_{\rm vac} = 0.73$) assuming a spectral index of zero, or a total luminosity of $1.1 \times 10^{45} \unit{erg~s^{-1}}$ over a bandwidth of 1~GHz. The rest-frame pulse width is  equal to about 7~ms (neglecting intergalactic scattering), implying a total energy release in the radio burst of $7 \times 10^{42} \unit{erg}$ over a bandwidth of 1~GHz. The energy in the radio burst fits well within the canonical $10^{51} \unit{erg}$ total energy release of the collapsar model, implying that the energy budget can support such a pulse. The pulse width of 7~ms implies a causative region of size $< 2.1 \times 10^8$~cm, some 10-100 times larger than the innermost stable orbit of a black hole, depending on its mass and spin, and well within the magnetosphere of a magnetar. In either case, this suggests that the emission emerges from a region close to the central engine.

Perhaps the two most confounding problems are the time delay between the gamma-ray trigger and the emission of the radio pulse, and that only one pulse is observed per GRB. If this is not an observational effect due to our detection threshold, the single detection implies that the radio pulse is related to a singular event that occurs 500-1000~s after the gamma-rays are released. If the central engine is a promptly formed black hole, it forms just before the release of gamma-rays. Therefore, there is no obvious mechanism related to the black hole, or GRB phenomenon, that can be invoked to explain the delay. On the other hand, for a magnetar central engine, \citet{Baumgarte00}  suggest that a super-Chandrasekhar ($>1.4 \unit{M_{\odot}}$) magnetar can be rotationally supported, and can avoid turning into a black hole as long as its angular momentum exceeds a threshold. If sufficient angular moment is lost (e.g. due to winds), a black hole is formed after some delay. \citet{Metzger11} suggest that the timescale for delayed black hole formation can be a few hundred seconds, but add that there is considerable uncertainty in this timescale. The exact mechanism for how this formation could release the radio pulse with the observed properties is still uncertain, but the delayed formation model has the desired characteristics: the delayed black hole formation happens only once, and occurs some time after the GRB trigger.

The alignment of the radio pulse with breaks in the X-ray light curve could also be related to the delayed formation of the black hole. The X-ray light curve is typically explained by a number of segments \citep[and references therein]{Gehrels09}: the initial rapidly falling afterglow is followed by a plateau lasting a few thousand seconds, which is in turn followed by a classical afterglow. The so-called plateau is usually explained as being due to additional energy being injected into the external shock, for example, from magnetar winds. The radio pulse after GRB 100704A occurs at the end of the X-ray plateau, which implies it is related to the end of the energy injection driving the plateau. One could imagine that the formation of the black hole event horizon encompassing the magnetar and its winds could be responsible both for the radio pulse and for the end of the X-ray plateau. On the other hand, the radio pulse following GRB 101011A occurs at the beginning of the X-ray plateau (which is not as clear as the plateau in GRB 100704A), casting some doubt on the delayed black hole formation as the source of this pulse.

The coincidence with the X-ray breaks could also be a statistical fluke. For GRB 101011A, the formal error in the break time is roughly 200~s, which is 10\% of the observation duration. There are at least two breaks in the X-ray light curve predicted during the observation interval of 200-1800~s \citep{Gehrels09}, suggesting a 20\% chance of a random event falling within the statistical errors of a break in the X-ray light curve. The constraints on the break time in GRB 100710A are less strong, implying an even higher chance of a random event falling within the errors of an estimated X-ray break.

\subsubsection{Not a GRB}
Given the spatial distribution of GRBs, we would expect to receive pulses with a range of flux densities, both near our threshold and far above it. The key argument against a GRB origin for these pulses is the small likelihood of our detections being so close to our detection threshold and the absence of any more significant detections.

To quantify this effect, we performed a monte-carlo simulation of $5\times 10^3$ GRBs. Our simulation assumed the GRBs were distributed uniformly in redshift from z=0 to 6 \citep{Jakobsson06}.  In the simulation we assume each  GRB releases a radio pulse of luminosity  $1.1 \times 10^{36} \unit{erg~s^{-1}~Hz^{-1}}$ (the luminosity derived in section \ref{sec:grb}). In practice, any emission mechanism is likely to have a more complex luminosity function, due to a range of factors such as initial stellar mass, environment and so on. As we have no model luminosity function, we fall back to the simplest case of a fixed luminosity. This luminosity function also reproduces our overall detection statistics. We assume that our detection threshold corresponds to a flux density of 7~Jy, and that our pulses were detected in the range 7 to 10~Jy. This range is somewhat larger than the actual range, but nonetheless represents a conservative interval.

The simulation implies we would detect pulses above our 7~Jy threshold in 26\% of cases (which compares well with our observed 25\%), but the probability of the flux density of a single pulse falling in the observed 7--10~Jy range is only 3\%. The simulation also implies that a single pulse would have flux density $> 10$~Jy in $23\%$ of cases. Furthermore, by applying the binomial theorem, the simulation implies we would obtain our observed statistics (2 pulses out of 8 GRBs in the range 7 to 10~Jy) with probability 2\%.

This simple simulation suggests that, if the pulses are of GRB origin, then the probability of detecting two such pulses so close to our detection threshold is small (2\%). We can say, therefore, that we have a 2\% confidence in a GRB origin for these pulses.

\subsubsection{Upper limits}

There remains a reasonable probability that neither of our pulses are associated with an astronomical phenomenon. In spite of our best efforts to prove otherwise, there could still be some factor we have not considered in our telescope configuration that could cause short, dispersed pulse in our data. The fact that the pulses are detected at relatively low significance (the brightest is at $6.6\sigma$), and that no theory has been proposed to explain the brightness or time delay of the pulse, implies that further observational and theoretical work is desirable to confirm such pulses as having a GRB origin.

We can say with certainty that we did not detect any pulses at a significance $>7\sigma$. Given our system equivalent flux density and processing method, we obtain  an upper limit on the detection of single dispersed radio pulses at 1.4~GHz between 200 and 1800s after a GRB of $1.27 w^{-1/2} \unit{Jy}$, where $6.4 \times 10^{-5} \unit{s} < w < 32 \times 10^{-3}  \unit{s}$ is the pulse width.

\subsubsection{Comparison with previous surveys}

Assuming our detections are real, our 7~Jy pulse with a spectral index of zero is consistent with the non-detection in the all-sky survey of \citet{Katz03} at 611~MHz, and also consistent with the non-detection of the follow-up observations of \citet{Koranyi95} \citet{Dessenne96} at 151~MHz.

Our data does not constrain the spectral index well, and if we are observing the same mechanism as the LB, the spectral index could be as steep as the LB ($\alpha = -4 \pm 1$). If we assume the LB spectral index, then our 7~Jy pulse is still consistent with a non-detection in the \citet{Katz03} survey, but if we scale our result to the parameters of \citet{Dessenne96} (the most sensitive of the 151~MHZ observations) our pulse would have a flux density of 57~kJy. The \citet{Dessenne96} observations do not exactly match ours, however. While the bandwidth smearing due to dispersion is negligible ($\sim 1$~ms for a bandwidth of 700~kHz), the lower time resolution of 1.5~s reduces their sensitivity to our 25~ms pulse from 16~Jy to 60~Jy. Nonetheless, our pulse should easily have been detected by their experiment, had they observed it. The non-detection of \citet{Dessenne96} can be explained if only a fraction of GRBs emit radio pulses. Alternatively, given that the LB spectral index is not very well constrained, a modification of the spectral index to $\alpha = -2.3$ is sufficient to put our pulse at the \citet{Dessenne96} detection threshold, after accounting for the lower time resolution. Therefore, it is possible to explain the \citet{Dessenne96} non-detections if radio pulses from GRBs have a somewhat flatter spectrum than measured for the LB.

Conservatively comparing our limits with the  all-sky result by \citet{Katz03} of 27~kJy, and assuming $\alpha = -4$ as for the LB, then our limit scales to $29.4 w^{-1/2} \unit{Jy}$ at 611~MHz, which is still substantially better. Comparing to the results of \citet{Dessenne96} at 151~MHz, our result limit scales to $1.3 \times 10^{4} w^{-1/2} \unit{Jy}$ assuming the worst case $\alpha = -4$ spectral index of the LB. The sensitivity of the \citet{Dessenne96} survey to pulses less than their time resolution of 1.5~s is $24 w^{-1} \unit{Jy}$ for $w < 1.5\unit{s}$ (ignoring scattering), so their result is more constraining for all time resolutions we probed.

\subsection{Long time integrations}

Ultimately, our experiment was not designed to detect flux changes over long time intervals. Our  feed was not equipped with a Dicke radiometer, which would have enabled us to calibrate the gain fluctuations (Section \ref{sec:long}) over the 1~s integration time, but would also have contaminated the high-time resolution studies. In addition, we did not regularly observe flux calibrators during our observations, as this would have ruined the periodicity searches (which require continuous data streams). Our  detection limits are therefore completely limited by the temperature-related gain variations in our system, of approximately $5\% \unit{hr^{-1}}$.  we would consider change of $20\% \unit{hr^{-1}}$ as a candidate, which implies that a change in source flux density of 760~Jy would have constituted a detection.

We did not detect any variation that was not immediately attributable to interference, either from satellites or the Sun. We set an upper limit of a change of 760~Jy on any long-duration emission ($>1$~s) from a GRBs within 200-1800~s, at 1.4~GHz.

\section{Conclusion}

We have searched for prompt radio emission from gamma-ray bursts at 1.4~GHz, using a robotic telescope and a pulsar backend. Our  telescope was typically on source within 200~s of the gamma-ray trigger.

We detected single dispersed pulses following two GRBs at significances $>6 \sigma$. Simple statistical arguments, and a null trial based on randomizing channels on existing data, rule out random fluctuations as the origin of these pulses at $95\%$ and $\sim 97\%$, respectively.  The arrival times of both pulses also coincide with breaks in the X-ray light curves. While high DM and absence of adjacent RFI lend credence to an astronomical origin, weak impulsive RFI and atmospheric origins of these pulses remain a distinct possibility. A simple population arguments suggests a GRB origin for these pulses of only 2\%. If the radio pulses are associated with the corresponding GRBs, they could be related to changes in the central engine, in particular  the delayed formation of a black hole due to spin-down of a rotationally-supported magnetar.

The non-detection of radio pulses by previous surveys of \citep{Koranyi95, Dessenne96, Katz03} can be explained by insufficient sensitivity of those surveys, a somewhat flatter spectrum than measured for the Lorimer burst, or the possibility that only a fraction of GRBs emit radio pulses.

If the single pulse is not related to the GRB, we set an upper limit on the flux density of radio pulses between 200 to 1800~s after GRB trigger of $1.27 w^{-1/2} \unit{Jy}$, where $6.4 \times 10^{-5} \unit{s} < w < 32 \times 10^{-3} \unit{s}$ is the pulse width. This limit is substantially better than the all-sky 27~kJy limit  of \citet{Katz03} at 611~MHz, although not as constraining as the limits on two GRBs by \citet{Dessenne96}.

We have detected no convincing repeating dispersed candidates. We also detect no candidates on timescales $>1 \unit{s}$, but our experiment was not primarily designed for such detections. Nonetheless, we set an upper limit of a change of 760~Jy on any long-duration emission ($>1$~s) between 200 to 1800~s from our GRB triggers.

The detection of single dispersed pulses in this experiment is intriguing. Clearly the next step is to determine whether these pulses are related to their GRBs, for which the key problem is ruling out RFI, statistical fluctuations and other equipment-related sources as the origin of these pulses. The simplest future experiment to rule out these origins is to use a coincidence detection, by employing the same telescope setup at two widely separated sites. The fact that some 25\% of GRBs may be accompanied by a radio pulse, even with our relatively poor sensitivity, suggests that sensitivity is not the key factor in this experiment. Therefore, similar dishes, feeds and backends can be used. More important parameters of this experiment are the short on-source time (preferably $<200$~s), and a wide separation between antennas. A simultaneous detection of a single pulse at two widely separated sites, even at $6 \sigma$, would almost certainly rule out RFI and statistical fluctuations, and render atmospheric effects a very remote possibility. If such a detection were to be made, the future would be wide open to probe the astrophysical and cosmological implications of these phenomena.

\acknowledgements
The authors would like to thank the HTRU team \citep{Keith10} for the loan of a spare IBOB, and Willem van Straten and Andrew Jameson for help with setting it up. We thank Sarah Burke-Spolaor for many enlightening discussions and help using the HTRU giant pulse pipeline, and Michael Keith for the idea of the null trial based on randomizing channels. We thank the members for the \emph{Swift} team supporting their excellent observatory. We would also like to thank the anonymous referee for their helpful comments. KB acknowledges the support of an Australian Postgraduate Award and a CSIRO top-up scholarship. The Centre for All-sky Astrophysics is an Australian Research Council Centre of Excellence, funded by grant CE11E0090. This research has made use of several facilities and tools including:

\begin{enumerate}

\item Data obtained through the High Energy Astrophysics Science Archive Research Center Online Service, provided by the NASA/Goddard Space Flight Center.

\item  Data obtained from the Leicester Database and Archive Service at the Department of Physics and Astronomy, Leicester University, UK

\item The Cosmology Calculator \citep{Wright06}.

\item \emph{HDF5 for Python}, A. Collette, 2008 \url{http://h5py.alfven.org}

\end{enumerate}

\bibliographystyle{apj}
\bibliography{Master.bib}

\end{document}